\documentclass{pnastwo}

\usepackage{graphicx}
\usepackage{txfonts}
\PassOptionsToPackage{hyphens}{url}
\usepackage{hyperref}
\usepackage{bm}
\usepackage{amssymb}

\DeclareMathAlphabet{\mathbfsf}{\encodingdefault}{\sfdefault}{bx}{n}

\def\apss{Astrophys. Space Sci.}
\newcommand{\apj}{ApJ}
\newcommand{\apjl}{ApJ}
\newcommand{\apjs}{ApJS}
\newcommand{\aap}{A \& A}

\newcommand{\mnras}{MNRAS}

\newcommand{\pre}{Phys Rev E}
\newcommand{\nat}{Nature}

\begin{document}

\title{Magnetic field evolution in magnetar crusts through three dimensional simulations}

\author{Konstantinos N. Gourgouliatos\affil{1}{Department of Applied Mathematics, University of Leeds, Leeds, LS2 9JT, UK}\affil{2}{k.n.gourgouliatos@leeds.ac.u},
Toby Wood\affil{3}{School of Mathematics and Statistics, Newcastle University, Newcastle upon Tyne, NE1 7RU, UK}
\and
Rainer Hollerbach\affil{1}{Department of Applied Mathematics, University of Leeds, Leeds, LS2 9JT, UK}
}

\contributor{Accepted by the Proceedings of the National Academy of Sciences of the United States of America}

\significancetext{The observed diversity of magnetars indicates that their magnetic topology is more complicated than a simple dipole. Current models of their radiative emission, based on axially symmetric simulations, require the presence of a concealed toroidal magnetic field having up to 100 times more energy than the observed dipole component, but the physical origin of such field is unclear. Our fully 3-D simulations of the crustal magnetic field demonstrate that magnetic instabilities operate under a range of plausible conditions, and generate small-scale field structures that are an order of magnitude stronger than the large-scale field. The Maxwell stress and Ohmic heating from these structures can explain magnetar bursts and surface hotspots, using comparable poloidal and toroidal magnetic fields.}

\maketitle

\begin{article}
\begin{abstract}
Current  models of magnetars require extremely strong magnetic fields to explain their observed quiescent and bursting emission, implying that the field strength within the star's outer crust is orders of magnitude larger than the dipole component inferred from spin-down measurements.
This presents a serious challenge to theories 
of magnetic field generation in a proto-neutron star. Here, we present detailed modelling of the evolution of the magnetic field in the crust of a neutron star through 3-D simulations.
We find that, in the plausible scenario
of equipartition of energy between
global-scale poloidal and toroidal magnetic components, magnetic 
instabilities transfer energy to non-axisymmetric, kilometre-sized magnetic features, in which the local field strength
can greatly exceed that of the global-scale field. These intense small-scale magnetic features can induce high energy bursts through local crust yielding, and the localised enhancement of Ohmic heating can power the star's
persistent emission. Thus, the observed diversity in magnetar behaviour can be explained with mixed poloidal-toroidal fields of comparable energies.
\end{abstract}


\dropcap{A}n estimate of the magnetic field intensity in a neutron star can be obtained by assuming that the observed spin-down is caused by electromagnetic radiation from a dipolar magnetic field \cite{Deutsch:1955}. Neutron stars for which this estimate exceeds the QED magnetic field  $4.4\times10^{13}$G are conventionally called magnetars, and typically exhibit highly energetic behaviour, as in the case of Anomalous X-ray Pulsars and Soft $\gamma$-ray Repeaters \cite{THOMPSON:1995,THOMPSON:1996}.
Puzzlingly, not all high magnetic field neutron stars exhibit energetic behaviour \cite{HABERL:2007}, and conversely, ``magnetar-like" activity has been observed in pulsars for which this field estimate is below \cite{REA:2010} or only marginally above the QED magnetic field \cite{AN:2013,SCHOLZ:2012,GAVRIIL:2002}. Furthermore, despite the high thermal conductivity of neutron stars' solid outer crusts \cite{POTEKHIN:1999},
observations of their thermal emission indicates that in some cases the surface is highly anisothermal \cite{GUILLOT:2015},
with kilometre-sized ``hot spots" thought to be produced by small-scale magnetic features \cite{Bernardini:2011}.
Phase resolved spectroscopy has revealed that some magnetars have small-scale magnetic fields whose strength exceeds their large-scale component by at least an order of magnitude \cite{TIENGO:2013,GUVER:2011},
which can be correlated 
with outbursting events \cite{RODRIGUEZ:2015}. These observations all imply that the magnetic field structure in magnetars is more complicated, and varied, than the traditional picture of a simple inclined dipole.

The origin of the extreme magnetic fields in these objects, which are the strongest found in nature, is uncertain.
Even if magnetic flux were exactly conserved during the star's formation, the resulting field would not exceed $10^{13}$G.  It seems likely then that the strong fields in magnetars must result from a combination of differential rotation and dynamo action prior to the formation of the crust \cite{SPRUIT:2008}.  Dynamo models generally predict magnetic fields with features over a wide range of scales, and poloidal and toroidal components of comparable strength \cite{Mosta:2015}.  However, once the crust forms, small-scale surface features in the magnetic field would
decay by Ohmic dissipation much faster than the global field.

The evolution of the crustal magnetic field is mediated by the Hall effect, corresponding to advection by free electrons \cite{GOLDREICH:1992}.
The Hall effect in magnetars typically operates on a shorter timescale than Ohmic dissipation, and might therefore explain the formation of small-scale magnetic features, whose dissipation could then power the star's thermal radiation.  However, axisymmetric simulations of the crustal magnetic field only generate such features if a toroidal magnetic field exceeding $10^{16}$G is assumed to reside within the crust \cite{PONS:2011, GEPPERT:2014}.  The toroidal field strength must significantly exceed that of the poloidal dipole, which is the only component that is measured from spin-down observations, presenting additional challenges to models of magnetic field generation.
A possible resolution is the growth of localised patches of strong magnetic field via magnetic instabilities. Such instabilities have been demonstrated in local, plane-parallel models \cite{RHEINHARDT:2004, WOOD:2014, GOURGOULIATOS:2015}, but not yet in a realistic global model.  

\section{Hall Evolution in Neutron Star Crusts}

Motivated by this puzzle we study the fully non-linear 3-D problem of the magnetic field evolution in a magnetar crust through numerical simulations, using a modified version \cite{WOOD:2015} of the PARODY code \cite{DORMY:1998,AUBERT:2008},
to determine the necessary conditions for the spontaneous generation of strong localised magnetic field out of a large-scale weaker one. The crust of the neutron star is treated as a solid Coulomb lattice, in which free electrons carry the electric current and the magnetic field evolution is described by Hall-MHD with subdominant Ohmic dissipation \cite{GOLDREICH:1992}. This process is described by the Hall-MHD induction equation:
\begin{eqnarray}
\frac{\partial \bm{B}}{\partial t} = -\nabla \times \left(\frac{c}{{4 \pi \rm e}n_{\rm e}} \left(\nabla \times \bm{B}\right)\times \bm{B} +\frac{c^{2}}{4 \pi \sigma} \nabla \times \bm{B}\right)\,,
\label{HALL}
\end{eqnarray}
where $\bm{B}$ is the magnetic induction, $n_{\rm e}$ the electron number density, $\sigma$ the electric conductivity, $c$ the speed of light and ${\rm e}$ the elementary charge. The first term in the right-hand-side of
equation~(\ref{HALL})  describes the Hall effect and the second one Ohmic dissipation. We assume a neutron star radius $R_{*}=10$km, and a crust thickness of $1$km. We shall express the quantities in terms of the normalised radial distance $r=R/R_{*}$, where $R$ is the distance from the centre. The electron number density is taken to be $n_{\rm e}=2.5 \times10^{34}$ cm$^{-3} \left(\frac{1.0463-r}{0.0463}\right)^{4}$, the electric conductivity is $\sigma=1.8\times 10^{23}$s$^{-1}  \left(\frac{1.0463-r}{0.0463}\right)^{8/3}$. Such profiles are good analytical fits of the form $\sigma \propto n_{\rm e}^{2/3}$ of more precise crust models \cite{CUMMING:2004} at temperatures $\approx 10^8\ {\rm K}$.

The code uses spherical harmonic expansions in latitude and longitude, and a discrete grid in radius. The linear Ohmic terms are evaluated using a Crank-Nicolson scheme, while for the non-linear Hall terms an Adams-Bashforth scheme is used.  We have used the Meissner superconducting condition assuming that no magnetic field penetrates the crust-core surface, serving as the inner boundary of the domain \cite{Hollerbach:2004}. This is a simplifying assumption, as the core may be a Type-II superconductor and magnetically coupled to the crust. Nevertheless, the exact structure of the field in the core is still uncertain with some works suggesting that the crustal magnetic field is practically disconnected from the core [c.f.~Figs.~3 and 4 of \cite{Henriksson:2013}, and Fig.~9 of \cite{Lander:2014}]. With respect to the outer boundary we matched the magnetic field to an external 
vacuum field satisfying $\nabla \times \bm{B}=\bm{0}$. Given the lack of any other 3-D code or analytical solution of the Hall evolution, the code has been benchmarked against axially symmetric calculations \cite{GOURGOULIATOS:2014b, GOURGOULIATOS:2014a} in various radial and angular resolutions with excellent quantitative agreement, and the free decay of Ohmic eigenmodes, in the absence of the Hall term, reproducing the analytical results. The results that we present here have a resolution of 128 grid-points in radius, and spherical harmonics $Y_\ell^m$ up to $\ell_{\rm max}, m_{\rm max}=60-80$. We have repeated some of the runs with the strongest magnetic fields, using twice the angular resolution, with $\ell_{\rm max}=m_{\rm max}=120$. We found good agreement in the overall magnetic field evolution of those runs with the lower resolution ones.

\section{Simulations: Initial Conditions and Results}

Given the complexity of the processes taking place during the formation of a neutron star, our knowledge of
its initial magnetic field 
is limited. We have therefore explored a wide range of initial conditions, based on three plausible scenarios regarding the magnetic field formation: the fossil field, the large-scale  dynamo, and the small-scale dynamo. In the scenario of a fossil field \cite{BRAITHWAITE:2004}, the magnetic field has relaxed dynamically to an axially symmetric twisted torus with comparable energy in its poloidal and toroidal components. The poloidal magnetic field has dipolar geometry and the toroidal magnetic field is strongest near the dipole equator; both components have $\ell=1$ symmetry, where $\ell$ is the spherical harmonic degree. The second scenario corresponds to a magnetic field formed through the combination of differential rotation and dynamo action in the proto-neutron star \cite{SPRUIT:2008, Mosta:2015}; in this case the most likely geometry is an $\ell=2$ toroidal and a dipolar poloidal field.  In these two scenarios the field is predominantly large-scale and axially symmetric. In the third scenario,
turbulent convection in the proto-neutron star
generates a small-scale dynamo magnetic field \cite{THOMPSON:2001}.
In this case, the field comprises small loops with size comparable to that of the convection cells.
\begin{figure}
\includegraphics[width=\columnwidth]{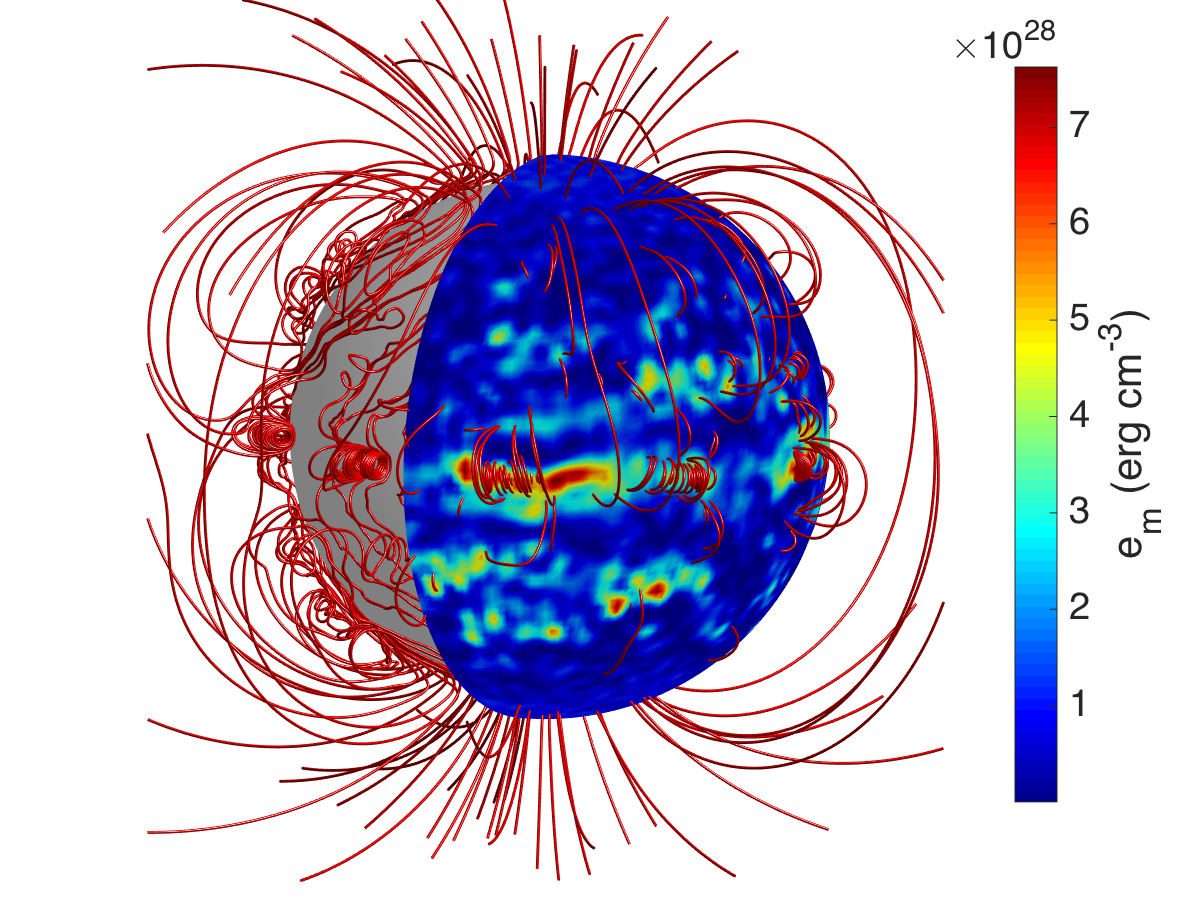}
\caption{Magnetic field lines in the simulation with an initially differentially twisted magnetic field consisting of an $\ell=1$ (dipole) poloidal and an $\ell=2$ toroidal field, starting with only $10^{-4}$ of the magnetic energy in the non-axisymmetric field, (run QU05-4, S.I.~Table 3). The snapshot is at $t=15$kyr overplotted with the magnetic energy density $e_{m}=B^{2}/(8\pi)$ on part of the surface. The surface field is highly anisotropic, with small regions in which the magnetic energy density exceeds by at least an order of magnitude the average surface value.
\label{Figure:1} }
\end{figure}

We have performed 76 simulations integrating the Hall-MHD induction equation
(\ref{HALL})
in each of the three scenarios for the initial magnetic field in the star's crust (see S.I.~Tables 1-3). In the fossil field and 
large-scale  dynamo scenarios we have used mixed poloidal and toroidal fields, varying the ratio of the toroidal to total magnetic energy and the overall normalisation of the magnetic field intensity. The intensity of the poloidal field on the surface takes values between $10^{13}$G to $5\times 10^{14}$G, except for the purely toroidal initial conditions.

The magnetic field is expressed in terms of two functions $V_p(r,\theta,\phi)$ and $V_{t}(r,\theta,\phi)$:
\begin{eqnarray}
\bm{B} = \nabla \times \nabla \times (V_p\bm{r})  + \nabla \times (V_{t} \bm{r})\,.
\end{eqnarray}
We express the fraction of the energy in the toroidal component of the magnetic field $e_{t}$, allowing the following values: $0$ (purely poloidal), $0.1$, $0.5$, $0.9$ and $1$ (purely toroidal). $B_{0}$ governs the overall amplitude and takes the values $0.5$, $1$, $2$ and $4$ expressed in units of $10^{14}$G; the normalisation is chosen so that the initial magnetic field energy for a simulation with amplitude $B_{0}$ is  $2.25\times 10^{46} B_{0}^{2}$ erg. For the radial structure of the field, we have used two basic profiles: one where the latitudinal and azimuthal components decrease monotonically with radius; and a second one with a local maximum in the middle of the crust. We present them below, providing the exact expressions to allow reproducibility of our results.

The first family of potentials corresponding to dipole fields ($\ell=1$), is given by: 
\begin{eqnarray}
V_{ps}&=&B_{0}(1-e_{t})^{1/2}~ \frac{3^{1/2}}{r^2}\cos\theta(-913.2837868+16749.59757r^3\nonumber  \\
&&-137700.2916 r^5 +292459.7609 r^6-269535.2857 r^7\nonumber \\
&&+120217.0825r^8-21277.28441r^9)\,,
\label{POTPS}
\end{eqnarray}
\begin{eqnarray}
V_{ts}=B_{0}e_{t}^{1/2}3^{1/2}~88.96339451 \cos\theta(r-1)\,.
\label{POTTS}
\end{eqnarray}
The poloidal part of this potential gives a dipole magnetic field supported by a uniformly rotating electron fluid  
corresponding to a Hall equilibrium \cite{GOURGOULIATOS:2013a}, with the magnetic field strength at the pole being equal to $B_{0}$ for $e_{t}=0$. Under this profile $B_{\theta}$ and $B_{\phi}$ change monotonically with radius. The maximum value of the $B_{\phi}$ component under this profile occurs at the base of the crust on the equatorial plane, and is $B_{\phi, {\rm max}}=15B_{0}e^{1/2}_{t}$. Thus, for the strongest toroidal field simulated ($B_{0}=4$, $e_{t}=1$), the maximum azimuthal field is $6\times 10^{15}G$.

The second family of $\ell=1$ profiles is the following:
\begin{eqnarray}
V_{pu}&=& B_{0}(1-e_{t})^{1/2}\frac{3^{1/2}}{r^{2}}\cos\theta (734.5987631-2333.649604r\nonumber \\
&&+2465.151852r^2-865.6887776r^3)\,,
\label{POTPU}
\end{eqnarray}
\begin{eqnarray}
V_{tu}=B_{0}e_{t}^{1/2}3^{1/2}~2739.401879 \cos\theta  (1-r)(r-0.9)\,.
\label{POTTU}
\end{eqnarray}
These profiles  have $B_{\theta}$ and $B_{\phi}$ components with an extremum close to the centre of the crust. 
The choice of normalisation is such that the magnetic energy in the crust is the same as the corresponding $V_{ps}$ and $V_{ts}$ profiles. The maximum value of the $B_{\phi}$ component under this profile occurs in the middle of the crust ($r=0.95$) on the equatorial plane and is $B_{\phi, {\rm max}}=12B_{0}e^{1/2}_{t}$. Thus, for the strongest toroidal field simulated ($B_{0}=4$, $e_{t}=1$), the azimuthal field has a maximum value of $4.8\times 10^{15}G$.

We also combined the poloidal dipole fields with
$\ell=2$ toroidal fields.
In such cases we have utilised the following profiles: 
\begin{eqnarray}
V_{ts,q}=\pm B_{0}e_{t}^{1/2}5^{1/2}~51.36303975 \left(\frac{3}{2}\cos^{2}\theta -\frac{1}{2}\right)(r-1)\,, 
\label{POTTSQ}
\end{eqnarray}
which is used in conjunction with $V_{ps}$, and 
\begin{eqnarray}
V_{tu,q}=\pm B_{0}e_{t}^{1/2}5^{1/2}~1581.594412 \left(\frac{3}{2}\cos^{2}\theta -\frac{1}{2}\right)(1-r)(r-0.9) \,,
\label{POTTUQ}
\end{eqnarray}
which is used in conjunction with $V_{pu}$. In the simulations with the quadrupolar toroidal field we assumed equipartition in energy between the poloidal and the toroidal field ($e_{t}=0.5$) and varied $B_{0}=0.5, 1, 2, 4$ for the case of the positive sign, while we run simulations with $B_{0}=2$ only for the case of the negative sign. The maximum value of the azimuthal field assumed, using the $V_{ts,q}$ profile is $5\times 10^{15}G$ and occurs at the base of the crust at latitudes $\theta=\pi/4, 3\pi/4$; while for the case of of $V_{tu,q}$ is $3.8\times 10^{15}G$ and occurs at $r=0.95$ and $\theta=\pi/4, 3\pi/4$. In all these axially symmetric initial conditions, we have superimposed a non-axisymmetric random noise, containing $10^{-4}-10^{-8}$ of the total energy. Axially symmetric initial conditions correspond to equilibria with respect to the $\phi$ coordinate, 
so any growth of the non-axisymmetric part of the field may indicate the presence of a non-axisymmetric instability.

Finally, we have simulated a highly non-axisymmetric magnetic field. In this case we have populated the crust with a fully confined non-axisymmetric magnetic field, where $\sim 0.9$ of the total magnetic energy is in the non-axisymmetric part of the field, consisting of multipoles with $10\leq\ell \leq 40$. Similarly to the previous cases we have used four normalisation levels for the overall magnetic field strength. 

We ran these simulations until the magnetic energy decayed to $0.01$ of its initial value. We also present a set of runs where the Hall term is switched off and the field evolves only via Ohmic decay.

Our basic conclusions are the following:
\begin{enumerate}
\item In systems where the initial azimuthal field is strong ($e_{t}\gtrsim 0.5$), and especially when the profiles $V_{pu}$ and $V_{tu}$ are used (Figs. 
\ref{Figure:1} and \ref{Figure:2}), the amount of energy in the non-axisymmetric part of the field increases exponentially. The growth rate is approximately proportional to the strength of the magnetic field (see S.I.~Figure 1), i.e.~for the case depicted in Figs.~\ref{Figure:1} and \ref{Figure:2} the non-axisymmetric energy growth timescale is $\sim 1.5$kyr. The exponential growth stops if one of the following two conditions is fulfilled: either $0.5-0.9$ of the total magnetic energy has decayed, or the non-axisymmetric field contains about 50\% of the total energy. While initially the $m=0$ mode was dominant, once the instability develops energy is transferred and a local maximum appears around $10\lesssim m \lesssim 20$ (see S.I.~Figure 3), with some further  local maxima at higher harmonics, leading to a characteristic wavelength for the instability of about $5$km at the equator, with finer structure appearing later giving kilometre-sized features. These features are well above the resolution of the simulation and appear both in the low and high resolution runs.  This behaviour is indicative of the density-shear instability in Hall-MHD, where, based on analytical arguments \cite{WOOD:2014, GOURGOULIATOS:2015}, the wavelength of the dominant mode is $2 \pi L$ where $L$ is the scale height of the electron density and the magnetic field, which can be approximated by the thickness of the crust. 

\begin{figure*}
\includegraphics[width=0.7\columnwidth]{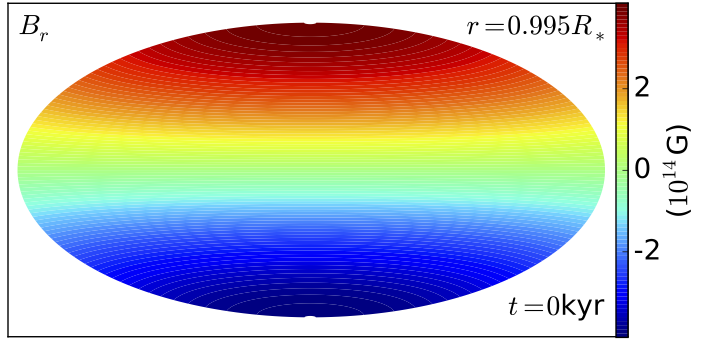}
\includegraphics[width=0.7\columnwidth]{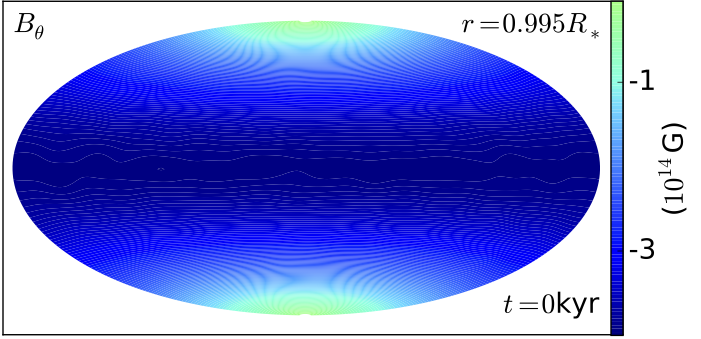}
\includegraphics[width=0.7\columnwidth]{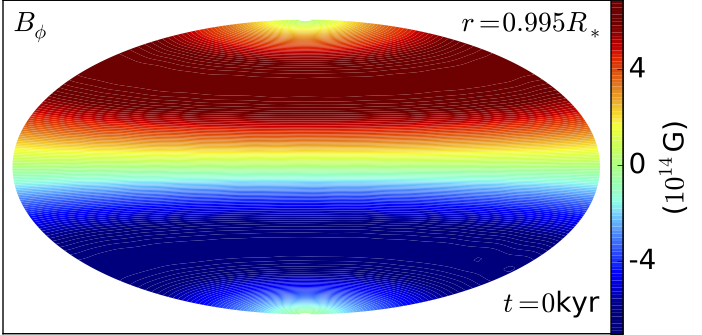}
\includegraphics[width=0.7\columnwidth]{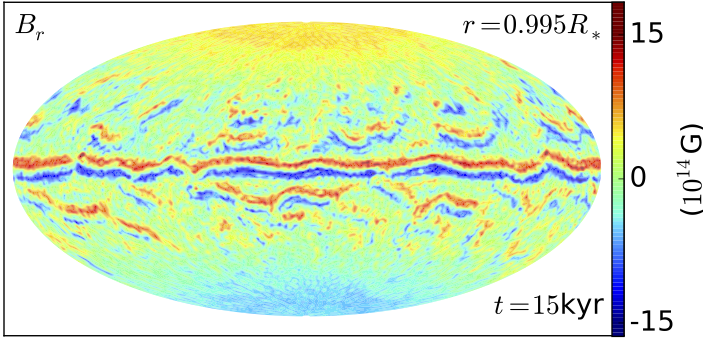}
\includegraphics[width=0.7\columnwidth]{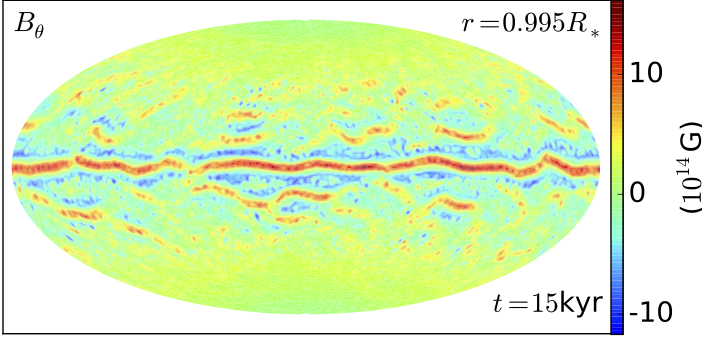}
\includegraphics[width=0.7\columnwidth]{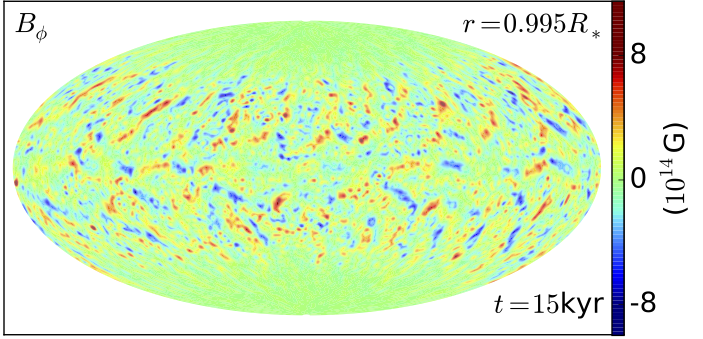}
\caption{Contour plot of the radial (left), latitudinal (middle) and azimuthal (right) component of the magnetic field at $r=0.995R_{*}$ at $t=0$ (top) and $t=15$kyr (bottom) for the simulation shown in Figure \ref{Figure:1}.
\label{Figure:2}}
\end{figure*}

\item  At the start of each simulation, the structure of the magnetic field changes rapidly, on the Hall timescale, unless the field is initiated in a Hall equilibrium in which case the evolution occurs on the slower Ohmic timescale. This change is accompanied by rapid magnetic field decay, i.e.~in the case of QU05-4 (S.I.~Table 3) 50\% of the energy decays within the first $20$kyr. We find that the early decay approximately scales with the magnetic field strength (Tables 1-3 S.I.~column $t_{0.5}$), whereas the later decay has a much weaker dependence on the magnetic field strength.
\item  Initial conditions with poloidal magnetic fields generate toroidal fields. The energy content of the toroidal field never becomes dominant irrespective of the strength of the initial field, (see S.I.~Figure 2). Even if a toroidal component exists in the initial conditions, stronger fields tend to suppress it more efficiently than weaker ones, in favour of the poloidal component. Thus a strong toroidal field is unlikely to be generated by a poloidal field being out of equilibrium. Once the magnetic field has decayed substantially the amount of the energy in the toroidal field may become higher than that of the poloidal. However, at this stage the dominant effect is Ohmic decay rather than Hall drift. This happens after a few Myrs as it is illustrated in the last column of the tables in the S.I.~for the simulations with $B_{0}=1$. 
\item The evolution of the axisymmetric part of the magnetic field is in accordance with the results of previous axisymmetric numerical and semi-analytical studies. In particular the toroidal field interacts with the poloidal field winding it up, while magnetic flux is transferred in the meridional direction leading to the formation of zones, Fig.~\ref{Figure:2}. The drift towards the equator is due to the polarity of the toroidal field, an effect that has been discussed in detail in plane-parallel geometry in \cite{Vainshtein:2000} and the axially symmetric case in \cite{Hollerbach:2004}.  
\item  Configurations initially dominated by the poloidal field, tend to remain axially symmetric, especially if the $V_{ps}$ profile is used. Their evolution is in general accordance with previous axially symmetric simulations that have described the interaction of the poloidal and toroidal part of the field. 
\item Once about 90\% of the magnetic energy has decayed the Hall effect saturates, and thereafter the magnetic energy decays on the Ohmic timescale.  However, the remaining small-scale features decay at the same rate as the large-scale field, indicating that the Hall effect is still active.  These results are indicative of the ``Hall attractor'' seen in axisymmetric simulations \cite{GOURGOULIATOS:2014b}. 
\item Initial conditions corresponding to highly non-axisymmetric structures undergo a weak inverse cascade (see runs Turb in the S.I.). The Hall effect transfers energy across the spectrum both to smaller and larger scales. Despite the great differences in the structure of the magnetic field compared to the axially symmetric case, the evolution follows a qualitatively similar pattern, with the rapid decay at early times, and saturation of the  Hall effect later.
\end{enumerate}

\section{Discussion}

In all simulations we find that the magnetic field undergoes a major restructuring at the very beginning of its evolution. This process occurs on the Hall timescale, and is thus faster for stronger magnetic fields. We then identified two distinct behaviours. Initial conditions where the poloidal field was dominant remain axially symmetric, while cases with substantial, but not necessarily dominant, azimuthal fields triggered non-axisymmetric instabilities. 
As magnetic energy is conserved in Hall-MHD, this growth of the non-axisymmetric field occurs at the expense of the axisymmetric components.

\begin{figure}
\includegraphics[width=\columnwidth]{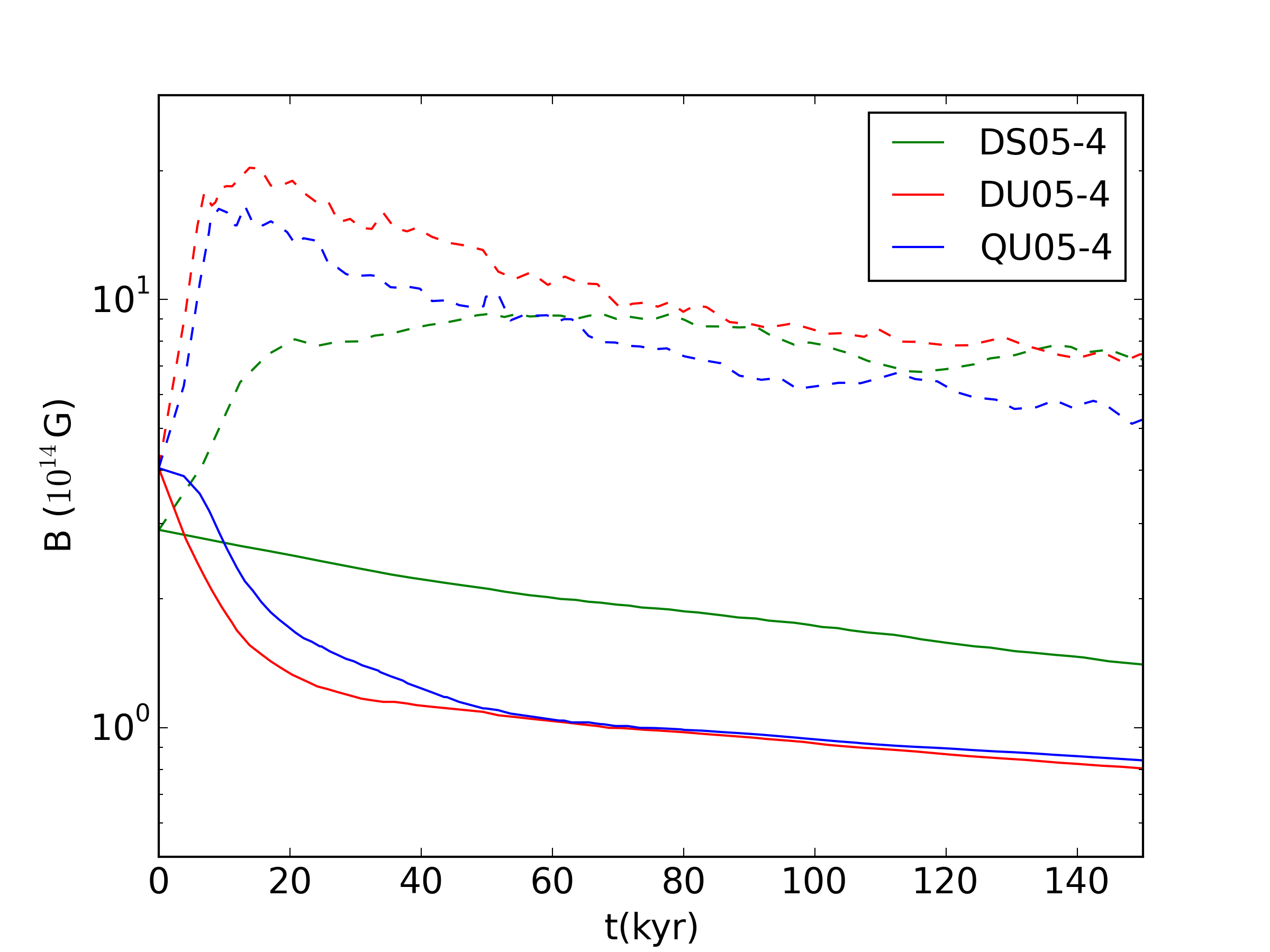}
\caption{The maximum magnetic field strength (dashed line) at the surface of the star versus the dipole field in three simulations. In the two cases where the non-axisymmetric instability is triggered (red and blue lines), the strongest field on the surface of the star exceeds the dipole component by an order of magnitude at a very early time, as opposed to the model that remains axially symmetric (green lines). 
\label{Figure:5}}
\end{figure}

\begin{figure}
\includegraphics[width=\columnwidth]{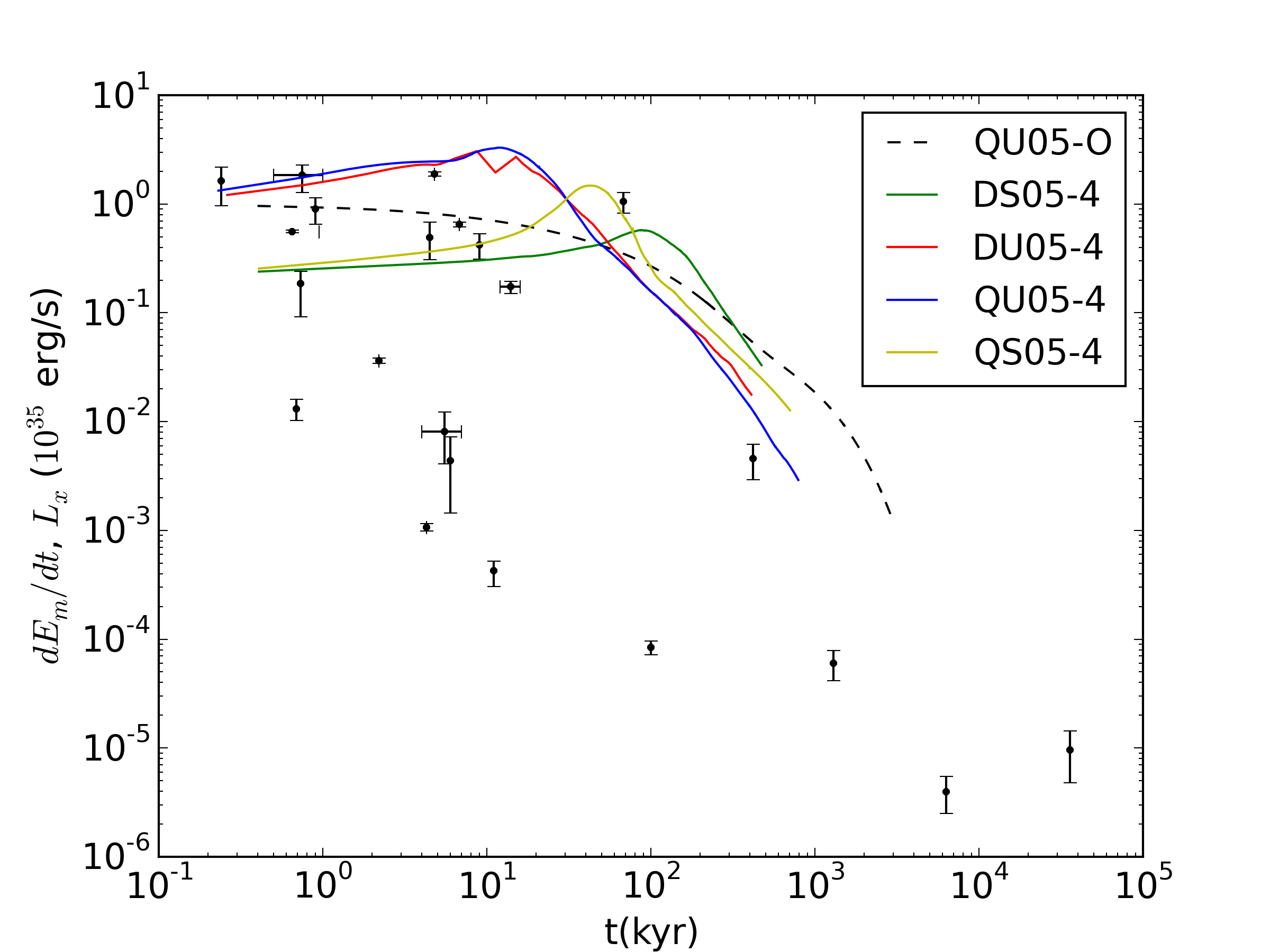}
\caption{The Ohmic heating rate versus time for the same simulations as Figure~\ref{Figure:5} (red, green and blue), each initially containing $3.6\times 10^{47}$erg of magnetic energy.  Also shown two simulations: one with an $\ell=1$ poloidal and an $\ell=2$ toroidal field (yellow line) where the latitudinal and azimuthal components decrease monotonically in the crust and the field remains mostly axisymmetric; and  a simulation where the structure of the field is identical to that of the blue line, in which the Hall effect has been artificially suppressed (dashed line). The dots represent observations of X-ray luminosities for magnetars in the McGill Magnetar Catalogue \cite{Olausen:2014}
plotted against their characteristic ages (or the age of the associated supernova remnant if available),
see S.I. Table 4. 
\label{Figure:4}}
\end{figure}

In the model with initial dipole poloidal magnetic field of $4\times 10^{14}$G and a toroidal component containing the same amount of energy with the poloidal (Figs.~\ref{Figure:1} and \ref{Figure:2}), magnetic spots of sizes $\sim 2$km appear. These spots contain energies up to $5\times 10^{43}$erg, while the dipole field  at that time is $1-2\times 10^{14}$G, (Fig.~\ref{Figure:5}). The rapid decrease of the dipole component, with a drop by a factor of 2 within 10-20 kyrs, is caused by the latitudinal Hall drift towards the equator. This is more efficient when the stronger fields are in shallow depths, when the potentials $V_{tu}$ and $V_{tu,q}$ are used (runs DU05-4 and QU05-4), where the maximum value of the toroidal field occurs at $r=0.95$. On the contrary it takes ten times longer for the same decrease in the dipole component, in the case of the potentials $V_{ts}$ (i.e.~run DS05-4), where the maximum of the toroidal field occurs at the base of the crust. Therefore, in the case of shallower fields the dipole field drops quickly, while there is still a large reservoir of energy deeper in the crust. This can be related to AXP 1E2259+586 and SGR 0418+5729 which despite their relative weak dipole field exhibit magnetar activity \cite{GAVRIIL:2002, REA:2010}. The energy in the magnetic spots is sufficient to power strong individual magnetar bursts or sequences of weaker events \cite{Mereghetti:2009, Horst:2012}. The energy in the equatorial zone of width $\sim~2$km is $\sim 2\times 10^{46}$erg, representing 20\% of the total magnetic energy in the crust at that moment. Such energy is comparable to the energy released by a magnetar giant flare \cite{Mazets:1979, Hurley:1999, Palmer:2005}. Note though, that magnetars that have exhibited giant flares have spin-down dipole fields above $5\times 10^{14}$G, so the overall magnetic energy should be scaled by a factor of $\sim10$ compared to our simulations. These structures take $\sim10^{4}$yrs to develop, given the choice of an initial condition where only $10^{-4}$ of the total is in the non-axisymmetric part. Had we chosen initial conditions with more energy in the non-axisymmetric part, the development of the smaller scale structure would have been imminent and in agreement with the ages of the most active magnetars, which are in the range of a few $10^{3}$yrs. This was confirmed in the extreme case where no energy is in the axisymmetric part (see runs Turb in S.I.~Table 3). Finally, we find that the Ohmic dissipation rate in these models provided enough thermal energy to cover the needs of young magnetars, (Fig.~\ref{Figure:4}), even though we have made a conservative choice of the magnetic dipole field.
By contrast, in cases where the evolution remains predominantly axisymmetric, instead of magnetic spots the field forms magnetic zones, where the intensity is only a factor of $\sim3$ stronger than the large-scale dipole field. Eventually, once a significant amount of energy has decayed ($\sim90\%$ of the total energy) the magnetic evolution saturates, and any small-scale magnetic features become ``frozen in'' as in previous studies in  Cartesian and axisymmetric simulations \cite{WAREING:2009b, GOURGOULIATOS:2014b}.

The diversity of observational manifestations of neutron stars has called for their Grand Unification \cite{Kaspi:2010} under a common theory, with their magnetic field being the key parameter \cite{Aguilera:2008b, Pons:2007, Pons:2009, Vigano:2013}. From our 3-D simulation survey we confirm indeed, that the initial structure and magnitude of the magnetic field are critical to the later evolution. We also find that the 3-D evolution is in general qualitatively different from the axisymmetric one. While predominantly poloidal initial magnetic field tend to remain axially symmetric \cite{WOOD:2015}, the inclusion of a strong, but not dominant, toroidal field is sufficient to break axial symmetry and has two major impacts, first: magnetic energy is converted faster to heat; 
second: small scale magnetic features of sizes appropriate for the hotspots observed in magnetars \cite{GUILLOT:2015} form spontaneously and persist for several $10^{4}$ yrs. This result is supportive of the importance of a second parameter, namely the toroidal magnetic field. However, contrary to  the axisymmetric studies which suggested that most of the energy is concealed in the toroidal field in order to explain bursts \cite{PONS:2011} and the formation of magnetic spots \cite{GEPPERT:2014}, a toroidal field containing the same amount of energy as the poloidal field is sufficient to power magnetars. This leads to a more economical magnetar theory compared to previous works. Moreover, it restores the theoretical consistency between the structure of the magnetic field in magnetars and the fact that the poloidal and toroidal fields inside a star need to be comparable \cite{Prendergast:1956,Flowers:1977} a result that has been confirmed numerically in MHD equilibria and dynamo studies \cite{BRAITHWAITE:2004,Mosta:2015}.
While the magnetic evolution in the crust is now well understood, it is critical for future studies to address fully the role of the core and its coupling to the crust, as the existing core studies focus on axisymmetric structures \cite{Lander:2013, Henriksson:2013, Elfritz:2015}. Furthermore, it is important to assess the impact of a magnetic field penetrating the crust-core boundary in the overall stability and Hall drift timescale. 
Similarly, it is possible that the strong magnetic field may plastically deform the crust, impeding the Hall evolution as the magnetic field lines are no longer frozen into the electron fluid \cite{Beloborodov:2014}. 
Regarding the thermal evolution, while it was shown here that the Ohmic decay provides sufficient energy to power magnetar X-ray Luminosity, it is important to couple magnetic and thermal evolution \cite{Vigano:2013} in a single 3-D calculation.  
Such a calculation will consider the possibilities of heat transfer deeper into the crust, suppression of radiation because of the non-radial magnetic field components, or losses to neutrinos. Finally other parameters such as the neutron star mass and crust thickness need to be taken into account in a wider exploration of physical mechanisms.

\section{Conclusions}

The observational implications of these results for magnetars are in three basic directions.
First, we have shown 
that the Hall effect has the tendency to produce
strong, small-scale magnetic fields.
This process does not appeal to an external source of energy, as the Hall effect exactly conserves magnetic energy, but rather to redistribution of the existing magnetic energy. 
The local magnetic field strength can
exceed $10^{15}$ G for a background dipole that is 
an order of magnitude weaker. Local magnetic fields of such strength are sufficient to reach the breaking stress \cite{ThompsonDuncan01,Lander:2015} and produce bursting activity.
Second, the enhanced Ohmic dissipation within the localised features generates sufficient heat to power their thermal X-ray luminosity. Therefore, the magnetic field will be used more efficiently to produce heat, making the overall energy requirements smaller, while the fact that the strong magnetic field is localised in smaller areas, gives a viable explanation for the existence of hotspots. Finally, the evolution of the magnetic field, under the Hall effect, saturates once the energy has decayed to $\sim 10\%$ of its initial value, within a few $10^{5}$yrs; in this case the neutron star still hosts a strong magnetic field $\sim 5 \times 10^{13}$G, but its slow magnetic evolution reduces the chances of bursts and flares. Therefore the most energetic behaviour occurs while the star is young, particularly for magnetars with very strong fields.

\begin{acknowledgments} 
KNG and RH were supported by STFC Grant No. ST/K000853/1. The numerical simulations were carried out on the STFC-funded DiRAC I UKMHD Science Consortia machine, hosted as part of and enabled through the ARC HPC resources and support team at the University of Leeds. We acknowledge the use of the McGill Magnetar Catalogue {\tt www.physics.mcgill.ca/$\sim$pulsar/magnetar/main.html}. We thank Andrew Cumming, Vicky Kaspi, Scott Olaussen, Yuri Levin, Anatoly Spitkovsky and Andrei Beloborodov for discussions and comments. We thank an anonymous referee for stimulating comments which improved the paper.

\end{acknowledgments}

\end{article}

\section{Supplementary Information}
We use the following naming convention: Simulations whose names start with DS correspond to combinations of $V_{ps}$ and $V_{ts}$ potentials, DU correspond to combinations of $V_{pu}$ and $V_{tu}$ potentials, QS correspond to combinations of $V_{ps}$ and $V_{ts,q}$ potentials,  QU correspond to combinations of $V_{pu}$ and $V_{tu,q}$ and those with Turb correspond to highly non-axisymmetric structures. The two numbers following the letters correspond to $e_{t}$, the ratio of the initial toroidal energy to the total energy, with 00 corresponding to purely poloidal field and 10 to purely toroidal. The number following the hyphen is related to the increments of the normalisation factor $B_{0}$, while simulations ending $-0$ correspond to runs where the Hall term is switched off and the evolution is purely Ohmic. Note that the simulations QS05-2M and QU05-2M utilise the negative sign for the potential $V_{ts,q}$ and $V_{tu,q}$ respectively. 

In the tables we provide the following information. The initial conditions are given in the first three columns (1-3); the following three columns (4-6) show the time it takes for the energy to decay to $0.5$ ($t_{0.5}$), $0.1$ ($t_{0.1}$) and $0.01$ ($t_{0.01}$) of its initial value; the rest of the table contains the values of the fraction of energy in the non-axisymmetric part of the field $E_{n}/E_{tot}$ (columns 7-10), and the ratio of the energy in the axisymmetric toroidal field over the total energy in the axisymmetric part of the field $E_{\phi}/E_{ax}$ (columns 11-14), at the instances determined by the decay levels mentioned before. Simulations appearing in Tables 1 and 2 start with setups corresponding to initial conditions of potentials $V_{ps}$, $V_{ts}$ (DS) and $V_{pu}$, $V_{tu}$ (DU) where the amount of energy in the non-axisymmetric part of the field is less than $10^{-7}$ of the total magnetic energy (except for the simulations that evolve only under the effect of Ohmic dissipation where the energy in the non-axisymmetric part of the field is $10^{-4}$ of the total energy). Simulations in Table 3 correspond to systems where the energy in the non-axisymmetric part is initially $\sim 10^{-4}$ of the total energy.

In the figures we illustrate the basic results of our simulations. In Fig.~\ref{Figure:Non-Ax} we plot the fraction of the energy in the non-axisymmetric part of the magnetic field. We find that (with the exception of the purely poloidal initial condition DS00-3) in all other cases the energy in the non-axisymmetric part grows exponentially.  In Fig.~\ref{Figure:Tor} we plot the fraction of the energy in the axisymmetric toroidal component for various runs, showing that a poloidal field does not generate a dominant toroidal field, and it actually even suppresses the initial toroidal field. In Figs.~(\ref{Figure:Ylm_Q}-\ref{Figure:Ylm_Turb}) we plot the energy decomposition in $\ell$ and $m$ modes of the spherical harmonics for various runs.

\setcounter{figure}{0}
\renewcommand{\thefigure}{S\arabic{figure}}
\begin{figure}
\includegraphics[width=0.5\columnwidth]{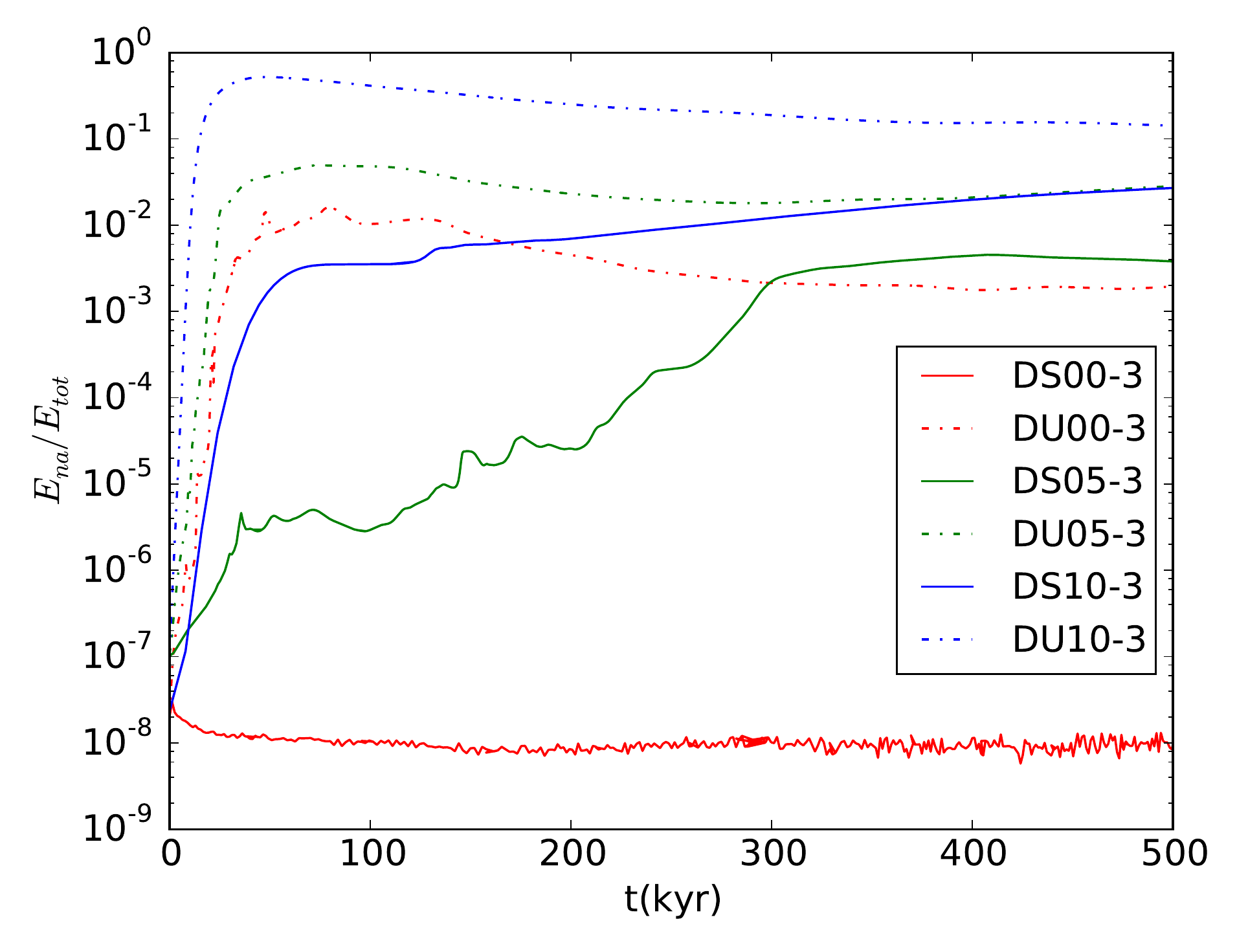}
\caption{Ratio of the energy in the non-axisymmetric part of the field over the total magnetic energy for six runs. In all runs except for the DS00-3, which corresponds to a poloidal Hall equilibrium supported by a uniformly rotating electron fluid, the amount of energy in the non-axisymmetric part of the field grows exponentially. The exponential growth is considerably faster when $V_{pu}-V_{tu}$ potentials are utilised (dotted lines) compared to $V_{ps}-V_{ts}$ and when the amount of energy in the toroidal field is larger.   }
\label{Figure:Non-Ax}
\end{figure}
\begin{figure}
\includegraphics[width=0.5\columnwidth]{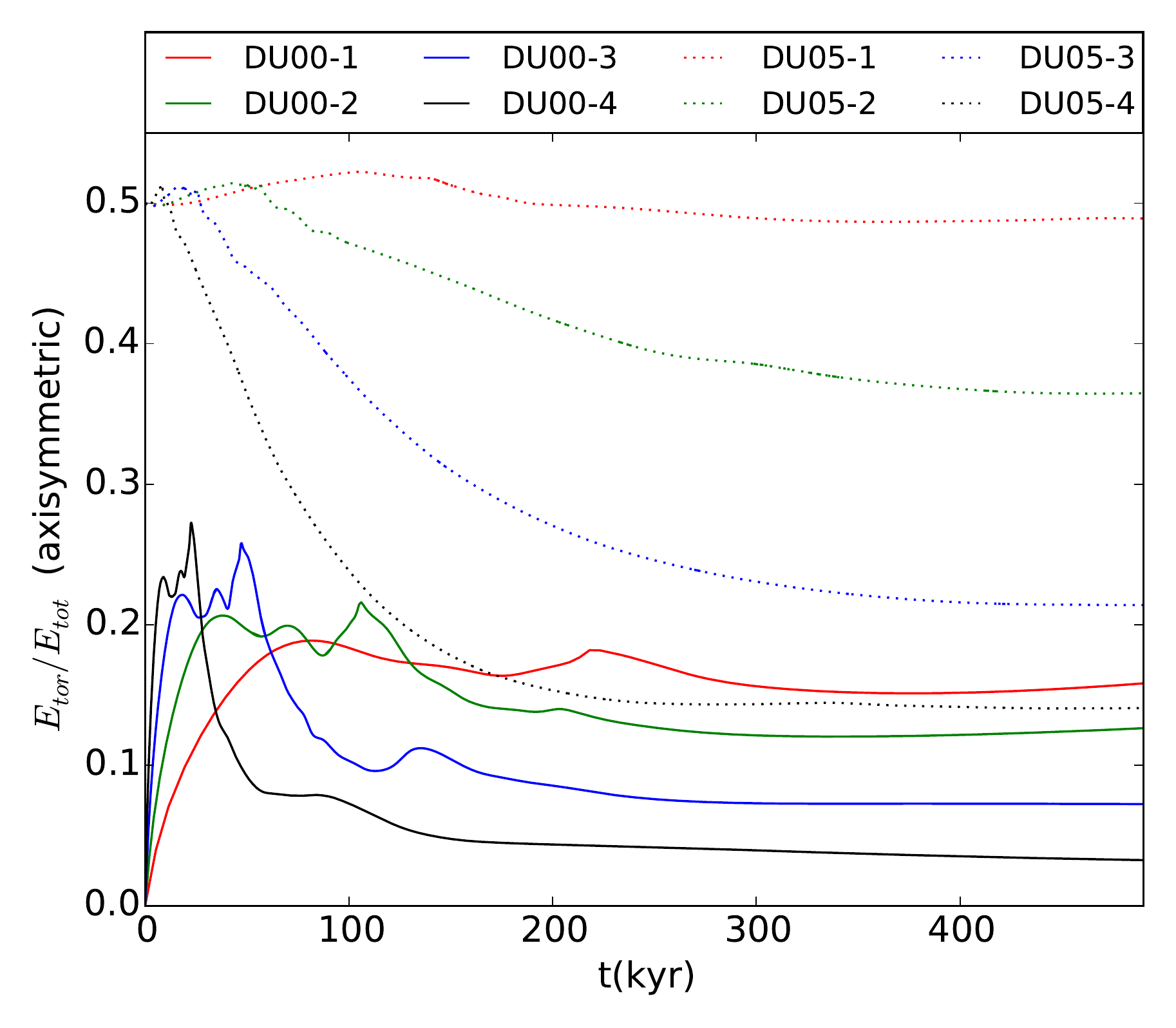}
\caption{Ratio of the energy in the axisymmetric toroidal component of the field over the total energy in the axisymmetric part of the field for eight different models. The DU00-1,2,3,4 runs start with a purely poloidal field, with intensities doubling between consecutive runs. The magnetic field configuration generates a toroidal component, as it is out of Hall equilibrium, however, the toroidal component does not become dominant, with stronger fields suppressing it efficiently and remaining effectively poloidal. Note that once a significant fraction of the field has decayed and the Ohmic term becomes strong, the fraction of energy in the toroidal field rebounds and increases (see for instance the red solid curve). Once the energy is equally split between the toroidal and the poloidal components, in the runs DU05-1,2,3,4 plotted with dotted lines, stronger fields again tend to suppress the toroidal field. }
\label{Figure:Tor}
\end{figure}
\begin{figure}
\includegraphics[width=0.5\columnwidth]{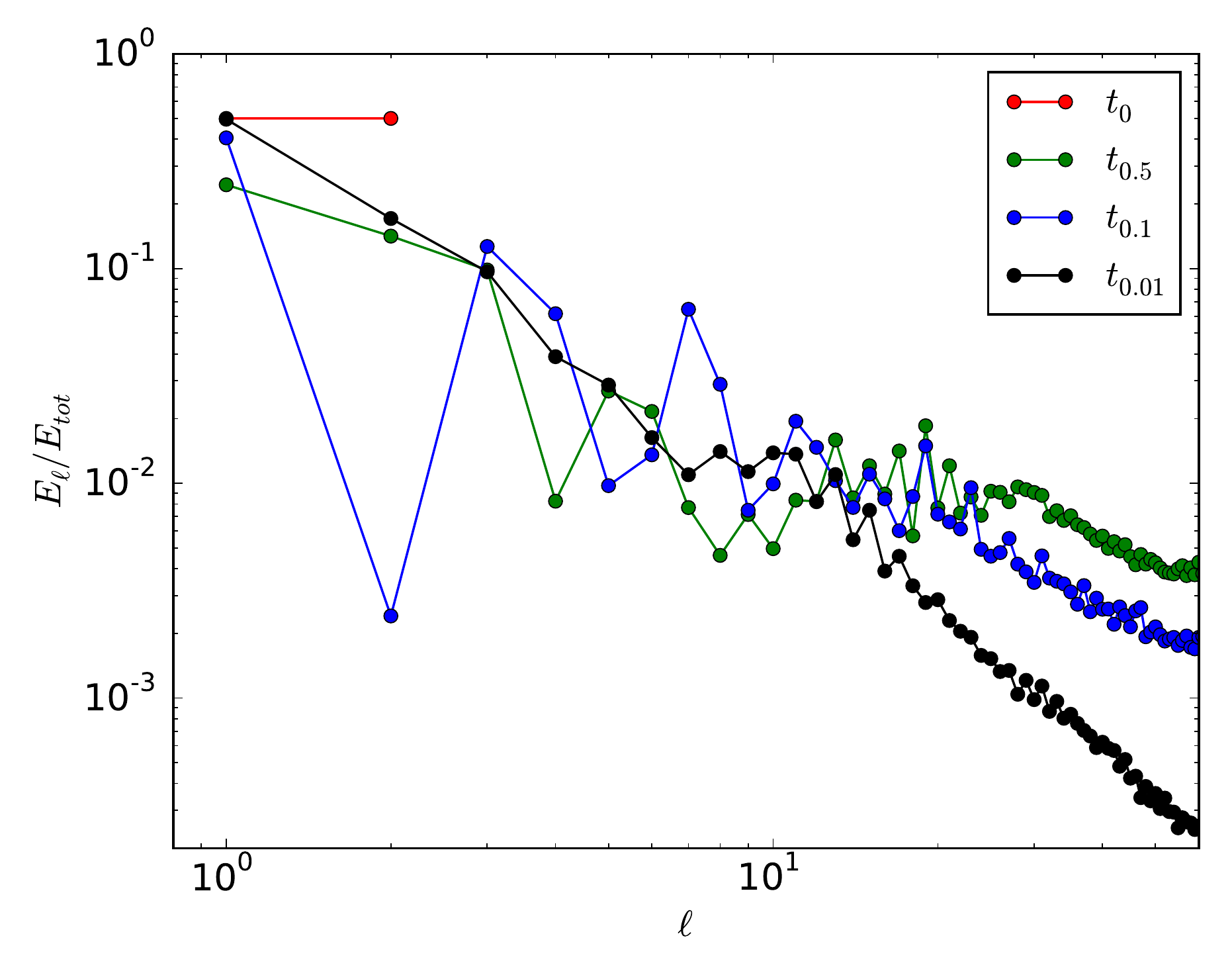}
\includegraphics[width=0.5\columnwidth]{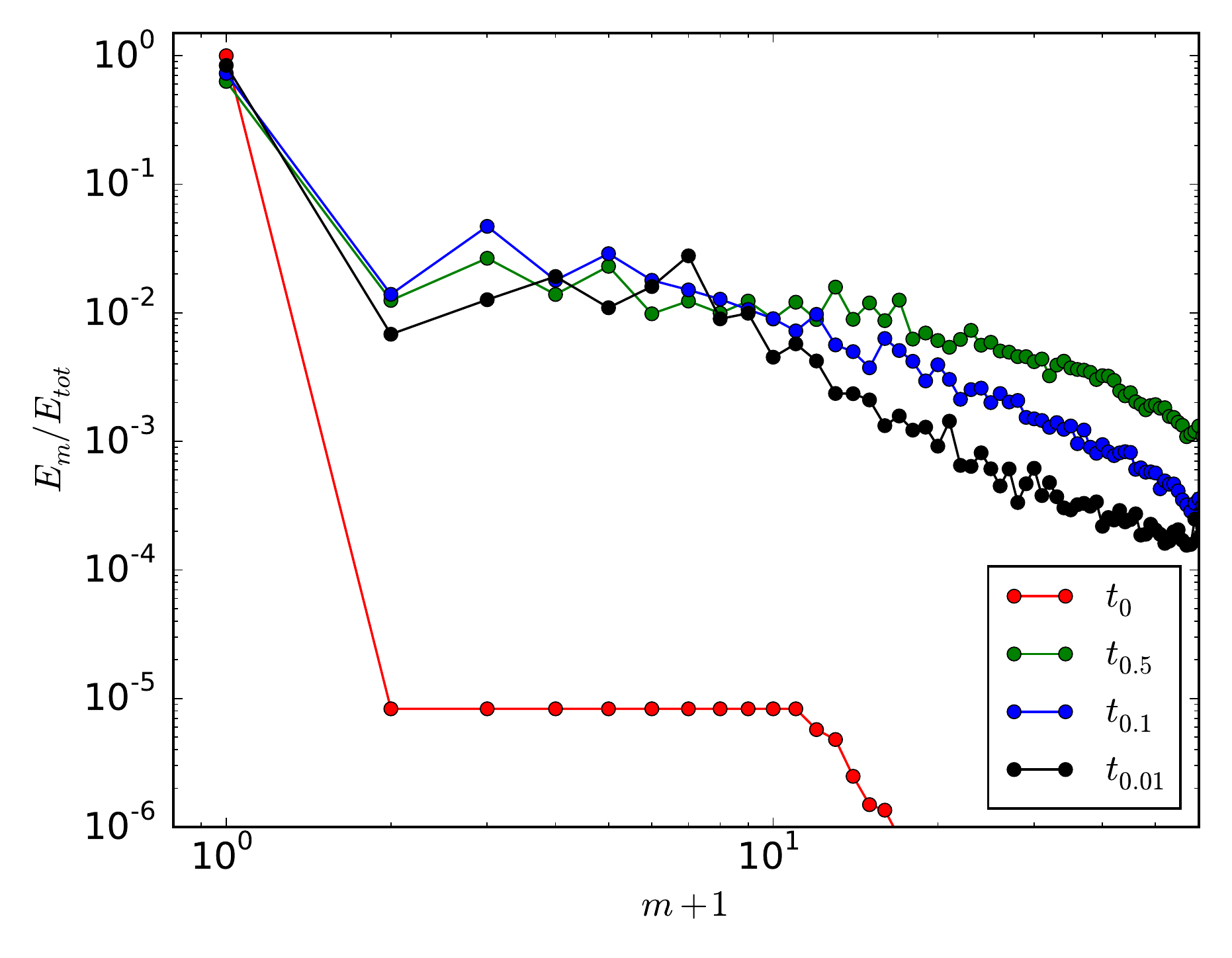}
\caption{Power spectrum of the $Y_{\ell}^{m}$ decomposition of the fields for run QU05-4, at times $t_{0}$, $t_{0.5}$, $t_{0.1}$ and $t_{0.01}$. Top panel: The $\ell$ spectrum, starting from a state where the energy is equally distributed in a poloidal $\ell=1$ field and a toroidal $\ell=2$ field. Early evolution excites higher modes with local maxima at characteristic scales, which are indicative of the zonal formations (please refer to Figure 3 of the main paper). Bottom panel: The $m$ spectrum; initially the energy is dominated by the axisymmetric mode $m=0$ and the system is in equilibrium in the azimuthal direction. Non-axisymmetric modes grow rapidly which is indicative of the unstable behaviour. }
\label{Figure:Ylm_Q}
\end{figure}
\begin{figure}
\includegraphics[width=0.5\columnwidth]{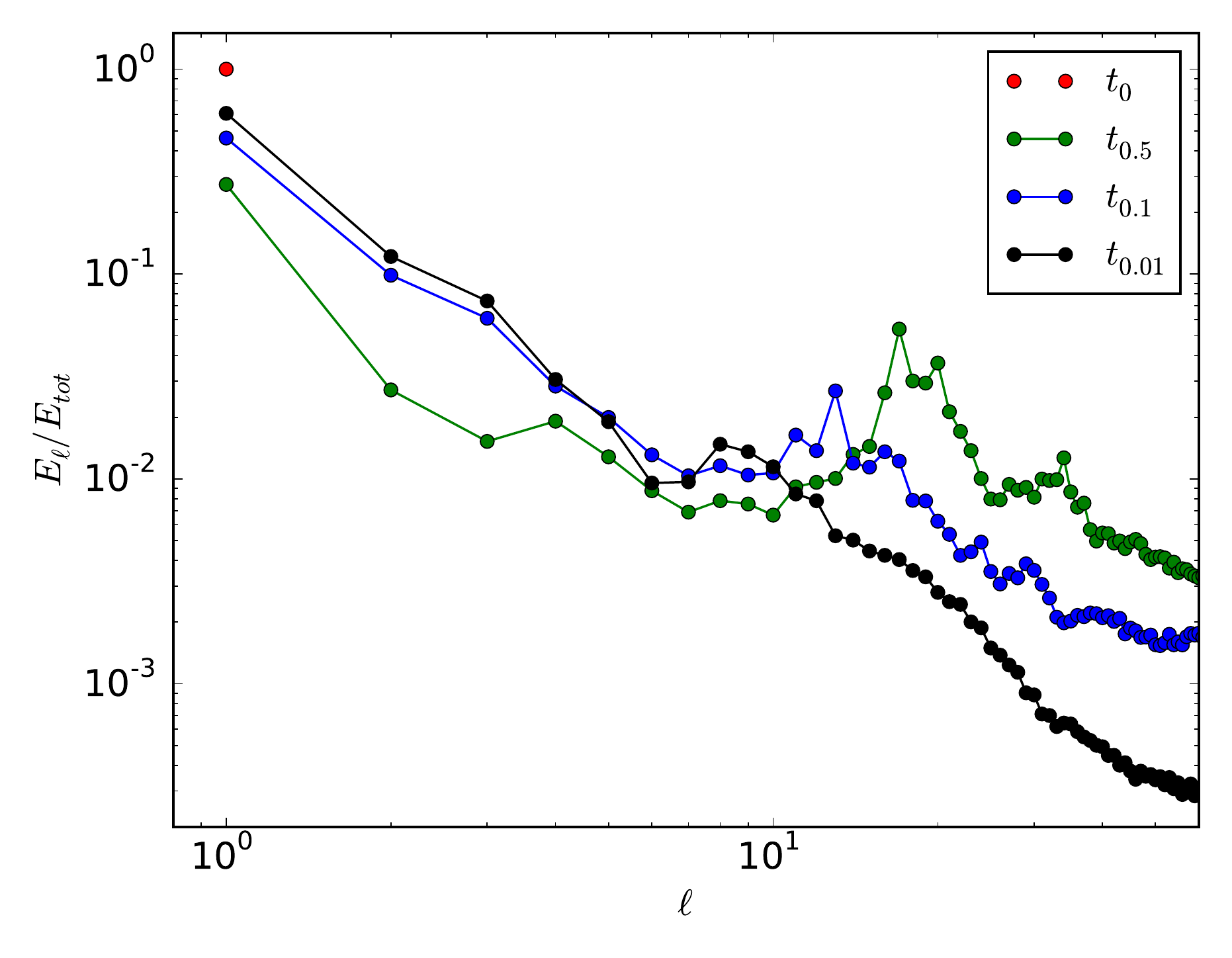}
\includegraphics[width=0.5\columnwidth]{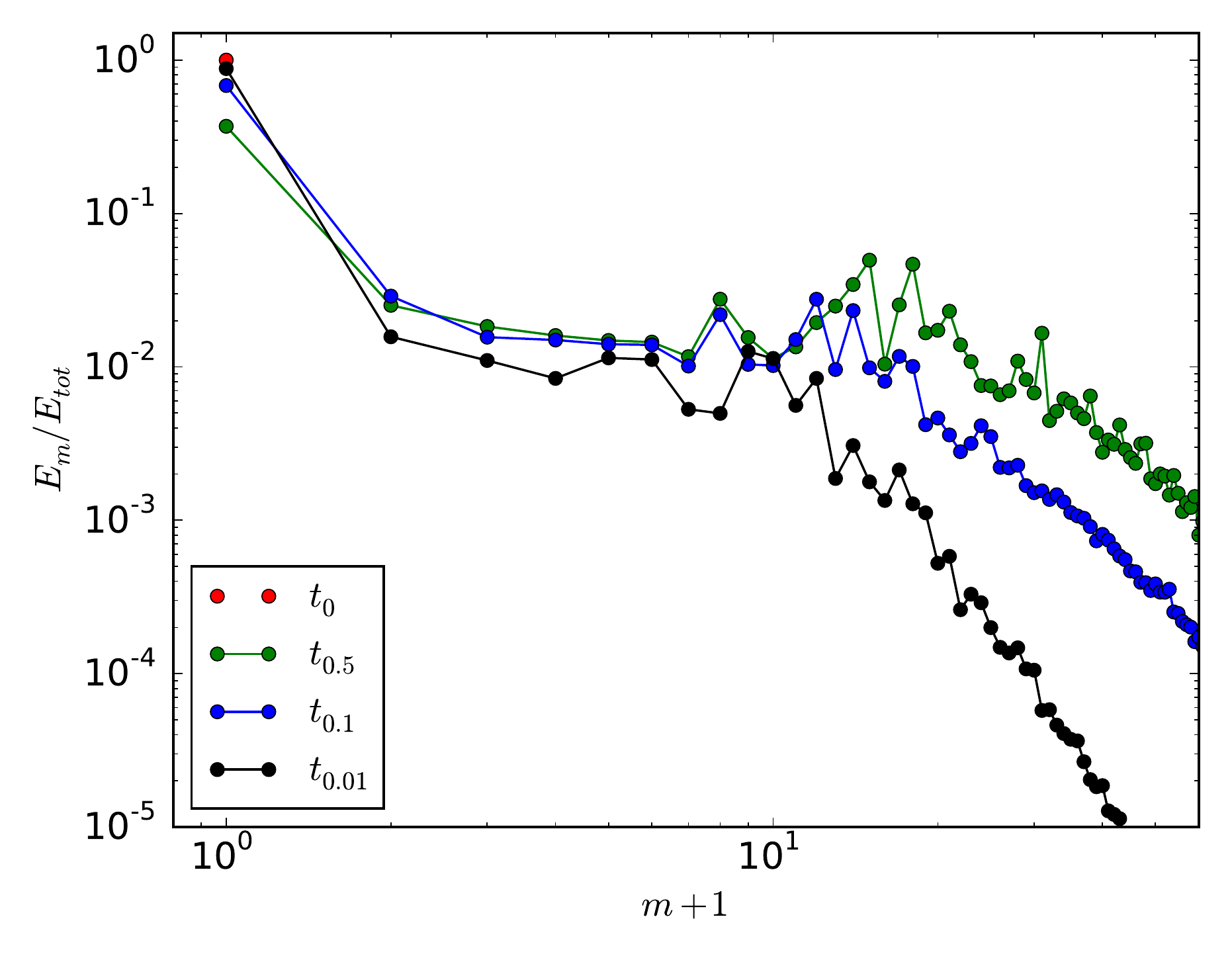}
\caption{Power spectrum as in Figure \ref{Figure:Ylm_Q} for run DU10-4. Top panel: The $\ell$ spectrum, starting from a state where all energy is contained in an $\ell=1$ toroidal mode. Hall evolution pushes energy into higher $\ell$'s especially early in the neutron star's life. Bottom panel: The $m$ spectrum, where initially the energy is dominated by the axially symmetric component $m=0$, however the development of Hall instability generates features with local maxima at $m=7,14,17,21,28,31$.}
\label{Figure:Ylm_Tor}
\end{figure}
\begin{figure}
\includegraphics[width=0.5\columnwidth]{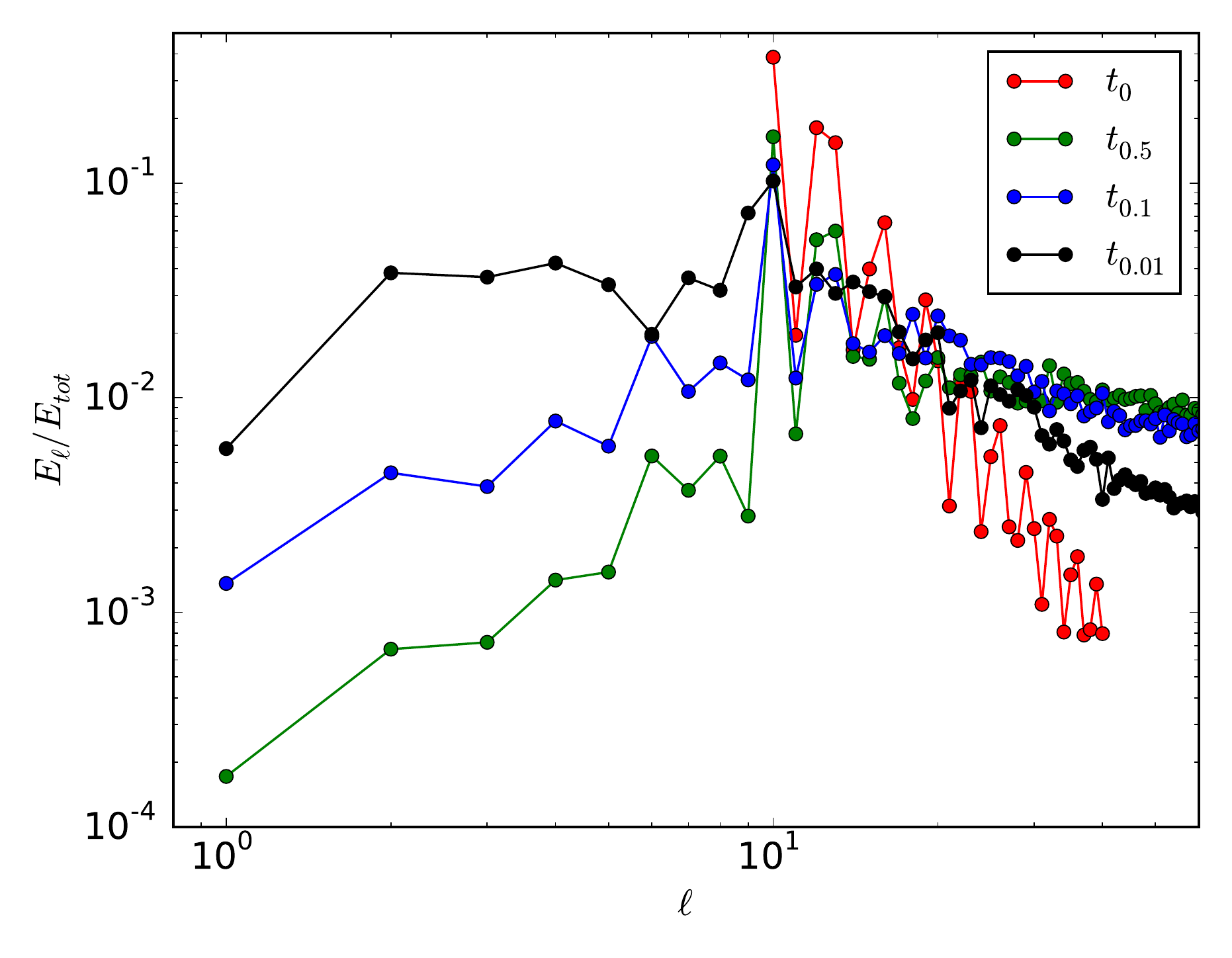}
\includegraphics[width=0.5\columnwidth]{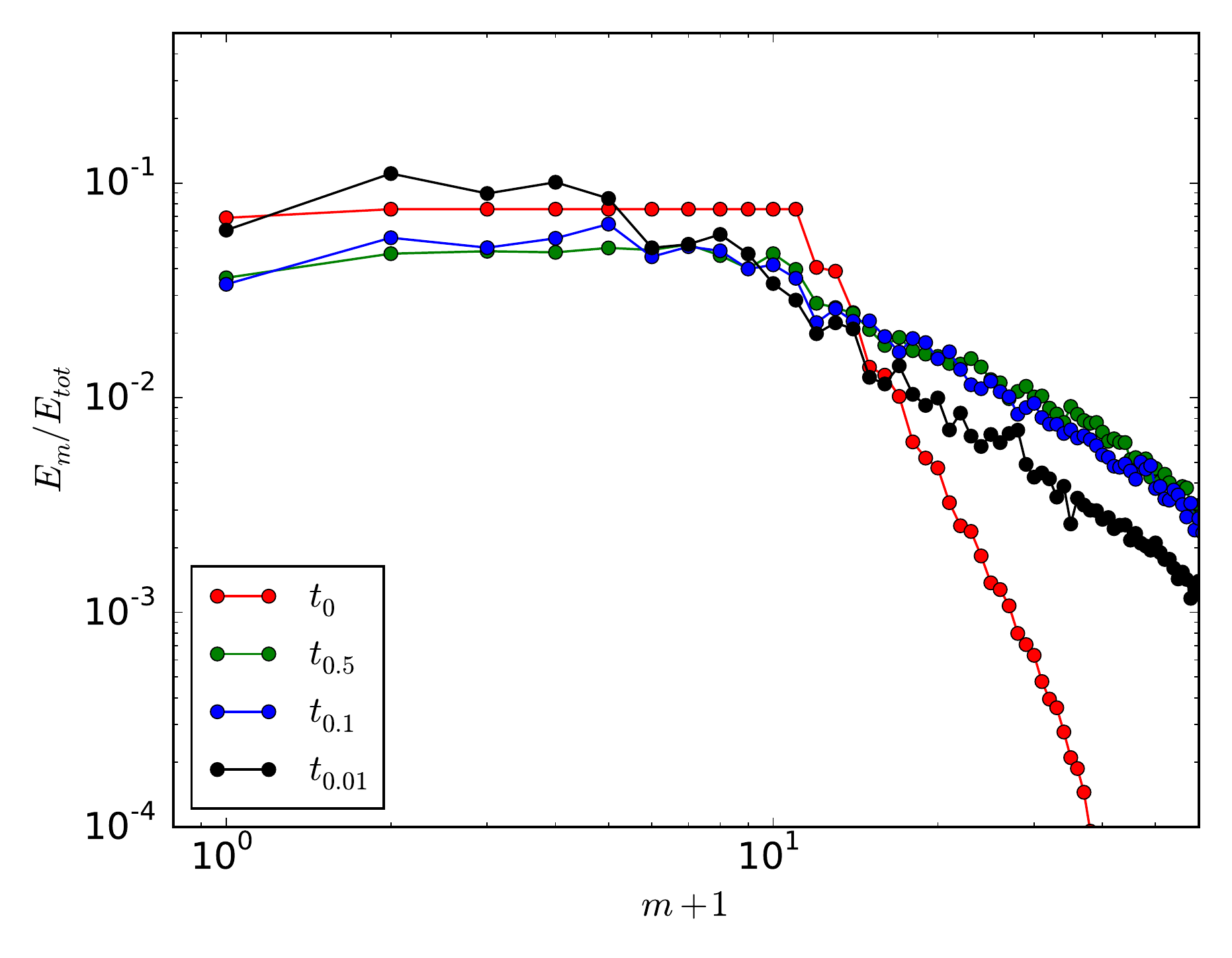}
\caption{Power spectrum as in Figure \ref{Figure:Ylm_Q} for run Turb-4. Top panel: The $\ell$ spectrum, starting from a state where the energy is distributed between $10\leq\ell \leq 40$. Energy is transferred both to smaller and larger scales, even to $\ell=1$, with the fraction of energy in this mode growing with time. Bottom panel: The $m$ spectrum, initially the energy is set to be equally distributed between $0\leq m \leq 10$ and then drops obeying a power-law. }
\label{Figure:Ylm_Turb}
\end{figure}

\begin{table}[h]
\caption{Simulations corresponding to $V_{ps}$ and $V_{ts}$ initial conditions. Times are expresses in $10^{3}$years. } 
\centering
\begin{tabular}{l c c crrrrrrrrrr} \hline\hline

NAME & $B_0$ & $e_t$ & $t_{0.5}$ & $t_{0.1}$ & $t_{0.01}$ & $E_{n}/E_{tot}(t_0)$ & $E_{n}/E_{tot}(t_{0.5})$ &$E_{n}/E_{tot}(t_{0.1})$ & $E_{n}/E_{tot}(t_{0.01})$ & $E_{\phi}/E_{ax} (t_{0})$ & $E_{\phi}/E_{ax} (t_{0.5})$ & $E_{\phi}/E_{ax} (t_{0.1})$ & $E_{\phi}/E_{ax} (t_{0.01})$ \\
\hline 
DS00-0 & - & 0 & 440 &  1585  &  3223  & 1.08E-04 & 2.28E-05 & 1.12E-05 & 6.05E-06 & 2.44E-06 & 6.56E-07 & 5.71E-07 & 5.73E-07 \\
DS00-1 & 0.5 & 0 & 376 &  1111  &  2176  & 3.70E-07 & 1.65E-07 & 1.30E-07 & 2.55E-07 & 6.40E-12 & 9.79E-02 & 2.38E-01 & 4.33E-01 \\
DS00-2 & 1 & 0 & 312 & 914 &  1841  & 9.57E-08 & 3.05E-08 & 7.06E-08 & 3.27E-07 & 0.00E+00 & 1.17E-01 & 1.44E-01 & 2.20E-01 \\
DS00-3 & 2 & 0 & 236 & 812 &  1733  & 1.99E-08 & 9.76E-09 & 2.92E-08 & 4.53E-07 & 4.55E-13 & 8.48E-02 & 5.01E-02 & 6.07E-02\\
DS00-4 & 4 & 0 & 187 & 783 & 1712 & 4.99E-09 & 2.94E-09 & 1.08E-08 & 3.75E-06 & 1.38E-13 & 4.33E-02 & 1.11E-02 & 1.21E-02 \\
\hline
DS01-0 & - & 0.1 & 448 & 1600  &  3249  & 1.08E-04 & 2.24E-05 & 1.09E-05 & 5.79E-06 & 9.98E-02 & 1.07E-01 & 1.17E-01 & 1.32E-01 \\
DS01-1 & 0.5 & 0.1 & 376 &  1111  &  2168  & 3.71E-07 & 2.07E-07 & 2.31E-07 & 5.04E-07 & 9.97E-02 & 1.53E-01 & 2.71E-01 & 4.73E-01\\
DS01-2 & 1 & 0.1 & 305 & 883 &  1797  & 9.30E-08 & 4.87E-08 & 2.25E-07 & 1.38E-06 & 9.97E-02 & 1.47E-01 & 1.70E-01 & 2.50E-01 \\
DS01-3 & 2 & 0.1 & 228 & 760 &  1672  & 2.49E-08 & 2.65E-08 & 2.81E-06 & 2.74E-06 & 9.97E-02 & 1.09E-01 & 6.77E-02 & 7.13E-02\\
DS01-4 & 4 & 0.1 & 171 & 730 & 1661 &  5.24E-09&  1.20E-05&  1.34E-04&1.80E-04   &  9.97E-02 &	6.74E-02&	1.63E-02&1.47E-02	  \\
\hline
DS05-0 & - & 0.5 & 471 &  1656  &  3351  & 1.00E-04 & 2.12E-05 & 9.83E-06 & 4.89E-06 & 5.00E-01 & 5.19E-01 & 5.45E-01 & 5.80E-01 \\
DS05-1 & 0.5 & 0.5 & 374 &  1088  &  2191  & 1.66E-06 & 6.89E-06 & 1.18E-05 & 2.03E-05 & 4.99E-01 & 4.23E-01 & 4.56E-01 & 6.93E-01 \\
DS05-2 & 1 & 0.5 & 287 & 773 &  1610  & 4.11E-07 & 2.46E-06 & 2.51E-05 & 1.33E-04 & 5.00E-01 & 3.28E-01 & 3.40E-01 & 4.81E-01 \\
DS05-3 & 2 & 0.5 & 197 & 558 &  1336  & 1.02E-07 & 2.53E-05 & 3.02E-03 & 2.17E-03 & 4.99E-01 & 2.42E-01 & 2.27E-01 & 1.94E-01\\
DS05-4 & 4 & 0.5 & 128 & 445 & 1305 & 2.50E-08 & 2.38E-03 & 7.45E-03 & 2.11E-02 & 4.99E-01 & 2.01E-01 & 1.07E-01 & 2.49E-02 \\
\hline
DS09-0 & - & 0.9 & 497 &  1705  &  3441  & 1.09E-04 & 2.00E-05 & 8.95E-06 & 4.20E-06 & 9.00E-01 & 9.07E-01 & 9.15E-01 & 9.26E-01 \\
DS09-1 & 0.5 & 0.9 & 420 &  1114  &  2458  & 3.72E-07 & 2.08E-05 & 2.69E-05 & 3.25E-05 & 9.00E-01 & 8.06E-01 & 8.32E-01 & 9.03E-01\\
DS09-2 & 1 & 0.9 & 289 & 717 &  1756  & 9.40E-08 & 1.21E-04 & 8.69E-04 & 1.73E-03 & 9.00E-01 & 6.94E-01 & 7.88E-01 & 8.96E-01 \\
DS09-3 & 2 & 0.9 & 169 & 417 &  1108  & 1.07E-07 & 1.85E-02 & 2.49E-02 & 9.51E-02 & 9.00E-01 & 6.35E-01 & 6.82E-01 & 8.23E-01\\
DS09-4 & 4 & 0.9 & 92 & 233 & 650 & 4.25E-09 & 5.33E-02 & 9.90E-02 & 2.16E-01 & 9.00E-01 & 5.87E-01 & 5.48E-01 & 6.27E-01 \\
\hline
DS10-0 & - & 1 & 497 &  1720  &  3464  & 1.09E-04 & 1.99E-05 & 8.73E-06 & 4.04E-06 & 1.00E+00 & 1.00E+00 & 1.00E+00 & 1.00E+00 \\
DS10-1 & 0.5 & 1 & 448 &  1175  &  2496  & 3.71E-07 & 9.03E-04 & 6.13E-04 & 2.34E-04 & 1.00E+00 & 1.00E+00 & 1.00E+00 & 1.00E+00\\
DS10-2 & 1 & 1 & 297 & 776 &  1841  & 9.27E-08 & 1.57E-03 & 4.35E-03 & 1.27E-02 & 1.00E+00 & 1.00E+00 & 1.00E+00 & 1.00E+00\\
DS10-3 & 2 & 1 & 169 & 461 &  1234  & 2.32E-08 & 6.26E-03 & 2.43E-02 & 6.21E-02 & 1.00E+00 & 1.00E+00 & 1.00E+00 & 1.00E+00 \\
DS10-4 & 4 & 1 & 92 & 259 & 794 & 4.55E-09 & 3.12E-02 & 5.61E-02 & 9.27E-02 & 1.00E+00 & 1.00E+00 & 9.99E-01 & 9.98E-01\\
\hline  \hline
\end{tabular}
\label{tab:LPer}
\end{table}

\begin{table}[h]
\caption{Simulations corresponding to $V_{pu}$ and $V_{tu}$ initial conditions, times expressed in $10^{3}$years. } 
\centering
\begin{tabular}{lcccrrrrrrrrrr} \hline\hline

\\ [0.5ex]

NAME & $B_0$ & $e_t$ & $t_{0.5}$ & $t_{0.1}$ & $t_{0.01}$ & $E_{n}/E_{tot}(t_0)$ & $E_{n}/E_{tot}(t_{0.5})$ &$E_{n}/E_{tot}(t_{0.1})$ & $E_{n}/E_{tot}(t_{0.01})$ & $E_{\phi}/E_{ax} (t_{0})$ & $E_{\phi}/E_{ax} (t_{0.5})$ & $E_{\phi}/E_{ax} (t_{0.1})$ & $E_{\phi}/E_{ax} (t_{0.01})$ \\
\hline\hline 
DU00-0 & 	- & 0 & 151 & 1096 & 2737 & 1.26E-04 & 6.83E-05 & 3.30E-05 & 1.63E-05 & 2.82E-06 & 2.19E-06 & 1.37E-06 & 1.17E-06 \\
DU00-1  & 0.5 & 0 & 90 & 637 & 1754 & 3.72E-07 & 2.37E-05 & 2.12E-05 & 1.14E-05 & 6.28E-10 & 1.88E-01 & 1.78E-01 & 3.69E-01 \\
DU00-2  & 1 & 0 & 67 & 530 & 1495 & 9.16E-08 & 9.89E-04 & 2.34E-04 & 3.16E-04 & 3.93E-10 & 1.98E-01 & 1.29E-01 & 2.32E-01  \\
DU00-3  & 2 & 0 & 46 & 474 & 1380 & 1.98E-08 & 1.32E-02 & 1.82E-03 & 1.04E-02 & 2.47E-10 & 2.58E-01 & 7.25E-02 & 9.30E-02 \\
DU00-4  & 4 & 0 & 28 & 404 &1206 & 7.34E-09 & 5.29E-02 & 3.58E-02 & 1.07E-02 & 1.56E-10 & 1.96E-01 & 3.51E-02 & 	3.31E-02\\
\hline
DU01-0 & 	- & 0.1 & 151 & 1103 & 2752 & 1.26E-04 & 6.83E-05 & 3.26E-05 & 1.58E-05 & 1.00E-01 & 1.00E-01 & 1.06E-01 & 1.20E-01 \\
DU01-1  & 0.5 & 0.1 & 90 & 648 & 1766 & 3.71E-07 & 2.14E-05 & 1.85E-05 & 1.30E-05 & 1.00E-01 & 2.49E-01 & 2.31E-01 & 4.26E-01 \\
DU01-2  & 1 & 0.1 & 67 & 530 & 1480 & 9.44E-08 & 6.17E-04 & 1.85E-04 & 2.31E-04 & 1.00E-01 & 2.45E-01 & 1.60E-01 & 2.64E-01 \\
DU01-3  & 2 & 0.1 & 46 & 443 & 1321 & 1.05E-07 & 3.16E-02 & 8.62E-03 & 3.24E-02 & 1.00E-01 & 2.27E-01 & 9.66E-02 & 1.05E-01 \\
DU01-4  & 4 & 0.1 & 28 & 351 & 1164 & 2.60E-08 & 8.90E-02 & 3.32E-02 & 	1.34E-01	 & 1.00E-01 & 2.02E-01 & 5.13E-02 &  3.77E-02  \\
\hline
DU05-0 & 	-	 & 0.5 & 151 & 1119 & 2816 & 1.26E-04 & 6.83E-05 & 3.14E-05 & 1.42E-05 & 5.00E-01 & 5.00E-01 & 5.17E-01 & 5.17E-01 \\
DU05-1  & 0.5 & 0.5 & 95 & 653 & 1869 & 3.71E-07 & 9.74E-05 & 1.10E-04 & 7.59E-05 & 5.00E-01 & 5.21E-01 & 4.95E-01 & 6.95E-01 \\
DU05-2  & 1 & 0.5 & 67 & 474 & 1418 & 9.20E-08 & 8.49E-03 & 4.01E-03 & 8.50E-03 & 5.00E-01 & 4.96E-01 & 3.65E-01 & 5.01E-01 \\
DU05-3  & 2 & 0.5 & 41 & 346 & 1119 & 1.04E-07 & 3.36E-02 & 1.98E-02 & 6.53E-02 & 5.00E-01 & 4.64E-01 & 2.22E-01 & 2.62E-01\\
DU05-4  & 4 & 0.5 & 23 & 236 & 887 & 2.50E-08 & 1.15E-01 & 8.52E-02 & 1.84E-02 & 5.00E-01 & 4.57E-01 & 1.46E-01 & 1.34E-01 \\
\hline
DU09-0 & 	- & 0.9 & 151 & 1144 & 2880 & 1.26E-04 & 6.83E-05 & 3.01E-05 & 1.27E-05 & 9.00E-01 & 9.00E-01 & 9.06E-01 & 9.18E-01 \\
DU09-1  & 0.5 & 0.9 & 118 & 809 & 2132 & 3.72E-07 & 1.50E-03 & 7.90E-03 & 6.24E-03 & 9.00E-01 & 8.67E-01 & 8.33E-01 & 9.03E-01 \\
DU09-2  & 1 & 0.9 & 82 & 517 & 1462 & 9.40E-08 & 1.97E-02 & 4.90E-02 & 4.35E-02 & 9.00E-01 & 8.46E-01 & 7.58E-01 & 8.97E-01 \\
DU09-3  & 2 & 0.9 & 51 & 292 & 909 & 1.04E-07 & 5.47E-02 & 1.09E-01 & 1.37E-01 & 9.00E-01 & 8.21E-01 & 6.55E-01 & 8.46E-01 \\
DU09-4  & 4 & 0.9 & 28 & 159 & 522 & 2.68E-08 & 7.88E-02 & 1.68E-01 & 2.98E-01 & 9.00E-01 & 8.01E-01 & 5.58E-01 & 7.32E-01 \\
\hline
DU10-0 & 	- & 1 & 151 & 1144 & 2895 & 1.26E-04 & 6.83E-05 & 2.99E-05 & 1.24E-05 & 1.00E+00 & 1.00E+00 & 1.00E+00 & 1.00E+00 \\
DU10-1  & 0.5 & 1 & 146 & 742 & 2086 & 3.73E-07 & 1.60E-01 & 5.38E-02 & 1.79E-02 & 1.00E+00 & 1.00E+00 & 9.99E-01 & 1.00E+00 \\
DU10-2  & 1 & 1 & 102 & 451 & 1454 & 9.27E-08 & 3.43E-01 & 1.25E-01 & 4.68E-02 & 1.00E+00 & 9.99E-01 & 9.99E-01 & 9.98E-01 \\
DU10-3  & 2 & 1 & 54 & 236 & 891 & 1.04E-07 & 5.16E-01 & 2.22E-01 & 1.09E-01 & 1.00E+00 & 9.95E-01 & 9.97E-01 & 9.93E-01 \\
DU10-4  & 4 & 1 & 28 & 123 & 530 & 2.70E-08 & 6.29E-01 & 3.14E-01 & 1.21E-01 & 1.00E+00 & 9.90E-01 & 9.83E-01 & 9.73E-01 \\
\hline  \hline
\end{tabular}
\label{tab:LPer}
\end{table}

\begin{table}[h]
\caption{Simulations starting with $10^{-4}$ of the total energy being in the non-axisymmetric component of the field, times expressed in $10^{3}$years. } 
\centering
\begin{tabular}{l c c crrrrrrrrrr} \hline\hline

\\ [0.5ex]

NAME & $B_0$ & $e_t$ & $t_{0.5}$ & $t_{0.1}$ & $t_{0.01}$ & $E_{n}/E_{tot}(t_0)$ & $E_{n}/E_{tot}(t_{0.5})$ &$E_{n}/E_{tot}(t_{0.1})$ & $E_{n}/E_{tot}(t_{0.01})$ & $E_{\phi}/E_{ax} (t_{0})$ & $E_{\phi}/E_{ax} (t_{0.5})$ & $E_{\phi}/E_{ax} (t_{0.1})$ & $E_{\phi}/E_{ax} (t_{0.01})$ \\
\hline\hline 
DS00-1  & 0.5 & 0 & 376 & 1111 & 2176 & 3.33E-04 & 1.50E-04 & 1.16E-04 & 2.28E-04 & 5.62E-09 & 9.79E-02 & 2.38E-01 & 	4.33E-01 \\
\hline
DS05-1  & 0.5 & 0.5 & 371 & 1085 & 2191 & 1.09E-04 & 3.56E-04 & 7.67E-04 & 1.21E-03 & 4.99E-01 & 4.23E-01 & 4.56E-01 &  6.93E-01\\ 
DS05-2  & 1 & 0.5 & 289 & 773 & 1608 & 1.08E-04 & 5.00E-04 & 4.46E-03 & 3.20E-02 & 5.00E-01 & 3.28E-01 & 3.40E-01 &  4.91E-01\\
DS05-3  & 2 & 0.5 & 197 & 553 & 1311 & 1.08E-04 & 1.53E-03 & 2.38E-02 & 1.48E-01 & 4.99E-01 & 2.41E-01 & 2.25E-01 &  2.20E-01\\ 
DS05-4  & 4 & 0.5 & 128 & 443 & 1250 & 1.09E-04 & 9.06E-03 & 2.37E-02 & 1.91E-01	 & 4.99E-01 & 1.98E-01 & 1.12E-01 &  2.74E-02\\	
\hline
DU05-1  & 0.5 & 0.5 & 95 & 655 & 1846 & 1.26E-04 & 1.43E-02 & 2.85E-02 & 2.27E-02 & 5.00E-01 & 5.20E-01 & 4.84E-01 &  6.77E-01\\ 
DU05-2  & 1 & 0.5 & 67 & 463 & 1388 & 1.26E-04 & 5.20E-02 & 5.09E-02 & 6.57E-02 & 5.00E-01 & 4.97E-01 & 3.66E-01 &  5.19E-01\\ 
DU05-3  & 2 & 0.5 & 44 & 323 & 1060 & 1.26E-04 & 1.09E-01 & 9.77E-02 & 1.36E-01 & 5.00E-01 & 4.68E-01 & 2.37E-01 &  3.02E-01\\
DU05-4  & 4 & 0.5 & 25 & 207 & 837 & 1.26E-04 & 1.76E-01 & 1.46E-01 & 2.43E-01 & 5.00E-01 & 4.42E-01 & 1.58E-01 & 	1.56E-01 \\
\hline
QS05-0  & 	- & 0.5 & 471 & 1649 & 3336 & 1.00E-04 & 2.00E-05 & 9.60E-06 & 4.91E-06 & 5.00E-01 & 5.18E-01 & 5.41E-01 & 5.73E-01\\	
QS05-1  & 0.5 & 0.5 & 256 & 783 & 1800 & 1.01E-04 & 1.71E-04 & 9.82E-04 & 1.44E-03 & 4.99E-01 & 5.63E-01 & 7.27E-01 &  9.76E-01\\
QS05-2  & 1 & 0.5 & 169 & 545 & 1288 & 1.00E-04 & 4.17E-04 & 2.56E-03 & 1.85E-02 & 5.00E-01 & 4.98E-01 & 5.12E-01 &  7.29E-01\\
QS05-3  & 2 & 0.5 & 102 & 417 & 1190 & 1.01E-04 & 1.20E-03 & 3.00E-03 & 6.86E-02 & 4.99E-01 & 4.08E-01 & 2.49E-01 &  2.17E-01\\
QS05-4  & 4 & 0.5 & 61 & 387 & 1267 & 1.01E-04 & 4.89E-03 & 3.89E-03 & 7.67E-02	  & 4.99E-01 & 3.19E-01 & 6.66E-02 & 	3.36E-02 \\  
\hline
QU05-0  & 	-	 & 0.5 & 151 & 1119 & 2808 & 1.00E-04 & 5.58E-05 & 2.61E-05 & 1.23E-05 & 5.00E-01 & 5.00E-01 & 5.14E-01 & 5.46E-01\\
QU05-1  & 0.5 & 0.5 & 84 & 571 & 1646 & 1.01E-04 & 4.68E-02 & 1.20E-01 & 1.02E-01 & 5.00E-01 & 2.37E-01 & 1.38E-01 &  1.86E-01\\ 
QU05-2  & 1 & 0.5 & 61 & 389 & 1226 & 1.00E-04 & 1.89E-01 & 1.69E-01 & 2.08E-01 & 5.00E-01 & 2.47E-01 & 1.21E-01 &  1.51E-01\\ 
QU05-3  & 2 & 0.5 & 38 & 259 & 952 & 1.01E-04 & 3.21E-01 & 2.23E-01 & 2.30E-01 & 5.00E-01 & 2.68E-01 & 1.12E-01 &  2.25E-01\\ 
QU05-4  & 4 & 0.5 & 20 & 161 & 709 & 1.01E-04 & 3.67E-01 & 2.74E-01 & 1.59E-01 & 5.00E-01 & 2.87E-01 & 1.24E-01 &  2.61E-01\\ 
\hline
QS05-2M  & 2 & 0.5 & 200 & 612 & 1449 & 1.01E-04 & 9.22E-03 & 2.09E-03 & 4.89E-03 & 4.99E-01 & 1.37E-01 & 1.83E-01 &  1.49E-01\\ 
QU05-2M  & 2 & 0.5 & 36 & 238 & 940 & 1.01E-04 & 5.38E-02 & 3.87E-02 & 2.23E-01 & 5.00E-01 & 5.59E-01 & 3.34E-01 &  3.28E-01\\
\hline
Turb-0  & 	- & 	- & 36 & 361 & 1231 & 9.31E-01 & 9.31E-01 & 9.27E-01 & 9.21E-01 & 	- & 3.27E-04 & 3.31E-04 & 	5.98E-04	\\ 
Turb-1 & 0.5 & 	-	 & 15 & 154 & 622 & 9.31E-01 & 9.48E-01 & 9.56E-01 & 9.52E-01 & 	- & 6.35E-02 & 6.62E-02 & 	6.62E-02\\ 
Turb-2 & 1 & 	-	 & 10 & 113 & 468 & 9.31E-01 & 9.53E-01 & 9.62E-01 & 9.48E-01 & 	- & 6.76E-02 & 9.05E-02 & 	8.28E-02\\				
Turb-3 & 2 & 	-	 & 8 & 77 & 335 & 9.31E-01 & 9.60E-01 & 9.65E-01 & 9.46E-01 & 	- & 7.81E-02 & 1.15E-01 & 	1.48E-01	\\ 
Turb-4 & 4 & 	-	 & 5 & 46 & 225 & 9.31E-01 & 9.65E-01 & 9.66E-01 & 9.39E-01 & 	- & 8.95E-02 & 1.11E-01 & 	 1.53E-01	\\
\hline  \hline
\end{tabular}
\label{tab:LPer}
\end{table}

\begin{table}[h]
\caption{Magnetar data used in Figure 5 of the main paper. The data are taken from the McGill Magnetar Catalogue (54). The error bars used in the plot are based in the distance uncertainties. With respect to age, the associated supernova remnant age is used in favour of the characteristic spin-down age $\tau_{c}$, if the former is known.  } 
\centering
\begin{tabular}{l c c c} \hline\hline
NAME & $L_{x}$ &$\tau_{c}$ (kyr) & SNR Age (kyr) \\
\hline 
CXOU J010043.1-721134&6.52E+34&6.8&-\\
4U 0142+61&1.05E+35& 68&-\\
SGR 0418+5729&9.57E+29& 36000&-\\
SGR 0501+4516&8.14E+32& 15& 4-7\\
SGR 0526-66&1.89E+35&3.4 & $\sim $4.8\\
1E 1048.1-5937&4.94E+34&4.5&-\\
1E 1547.0-5408&1.31E+33&0.69&-\\
PSR J1622-4950&4.36E+32&4& $<$6\\
SGR 1627-41&3.62E+33&2.2&-\\
CXOU J164710.2-455216&4.55E+32& 420&-\\
1RXS J170849.0-400910&4.20E+34& 9 &-\\
CXOU J171405.7-381031&5.59E+34&0.9& 0.65$^{+2.5}_{-0.3}$\\
SGR J1745-2900&1.07E+32& 4.3&-\\
SGR 1806-20&1.63E+35&0.24&-\\
XTE J1810-197&4.25E+31& 11&-\\
Swift J1822.3-1606&3.98E+29& 6300&-\\
Swift J1834.9-0846&8.44E+30& 4.9&$\sim$100\\
1E 1841-045&1.84E+35&4.6& 0.5-1\\
3XMM J185246.6+003317&6.03E+30& 1300&-\\
SGR 1900+14&8.97E+34&0.9&-\\
1E 2259+586&1.73E+34&230& 14(2)\\
PSR J1845-0258 &1.85E+34&0.73& -\\
\hline  \hline
\end{tabular}
\label{tab:LPer}
\end{table}

\end{document}